\documentclass[letterpaper]{article} 
\usepackage[preprint]{aaai2027}  
\usepackage[hyphens]{url}  
\usepackage{graphicx} 
\usepackage{natbib}  
\usepackage{caption} 
\usepackage{amsmath}
\usepackage{amssymb}
\usepackage{mathtools}
\usepackage{booktabs}
\usepackage{multirow}
\usepackage{colortbl}
\usepackage{xcolor}
\usepackage{enumitem}
\usepackage{microtype}

\graphicspath{{./figures/}}

\newcommand{\Description}[1]{}

\newcommand{\rkA}[1]{\cellcolor{red!30}#1}    
\newcommand{\rkB}[1]{\cellcolor{orange!30}#1} 
\newcommand{\rkC}[1]{\cellcolor{yellow!30}#1} 

\newcommand{\methodname}{BlitzGS}

\newcommand{\teaserfig}{%
  \vspace{-7em}
  \vskip 1.2em
  \begin{minipage}[c]{0.55\textwidth}%
    \centering
    \resizebox{0.99\textwidth}{!}{\includegraphics{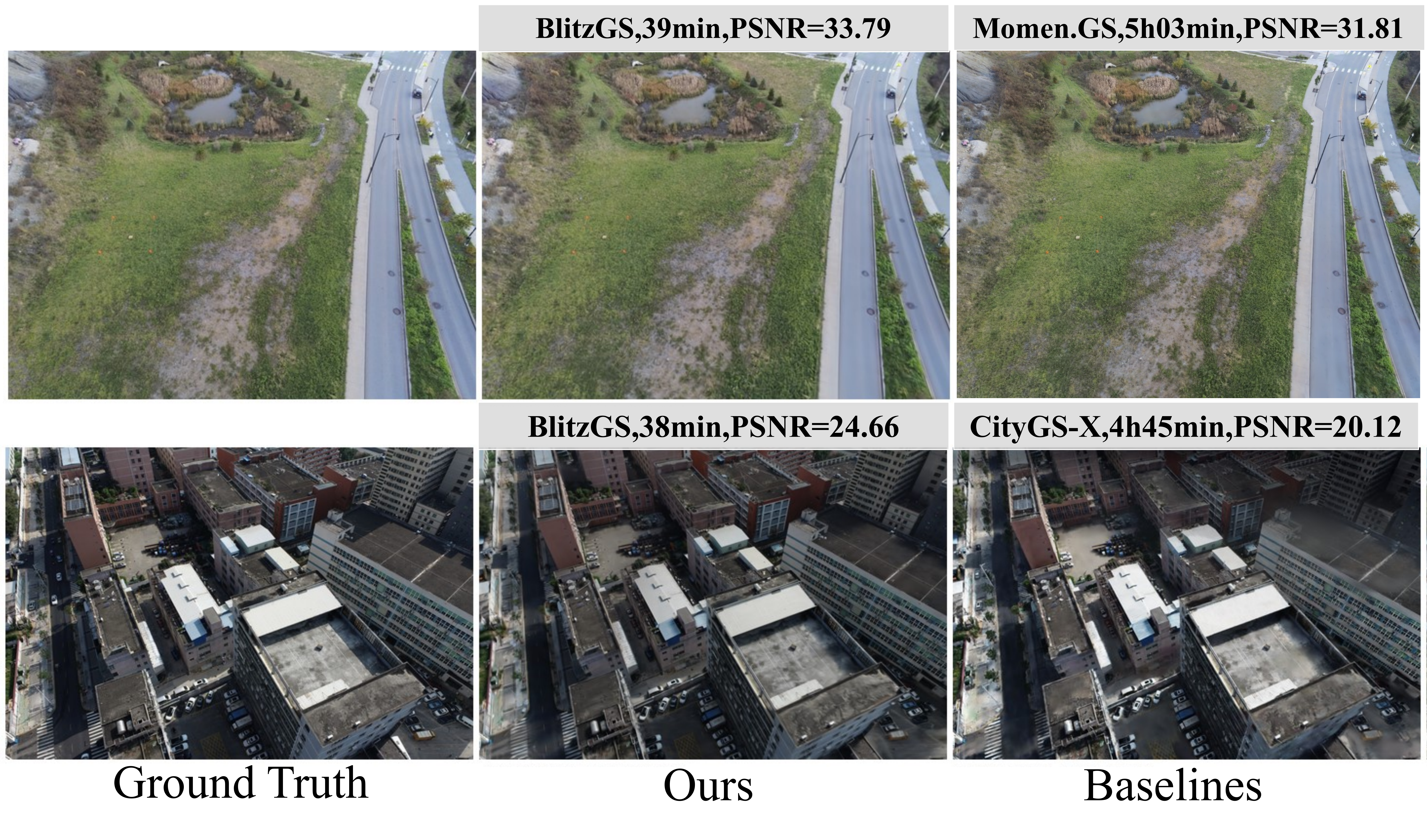}}%
  \end{minipage}%
  \hfill
  \begin{minipage}[c]{0.44\textwidth}%
    \centering
    \resizebox{0.99\textwidth}{!}{\includegraphics{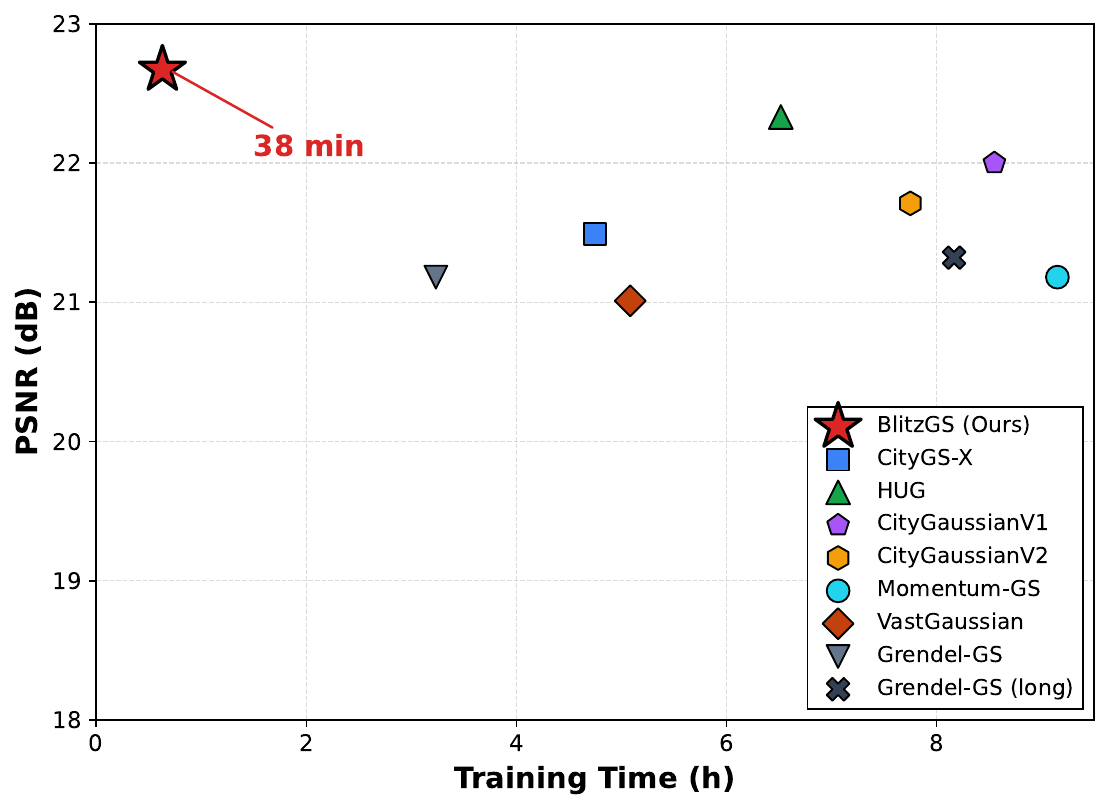}}%
  \end{minipage}%
  \vskip 0.4em
  \vspace{-1em}
  \captionof{figure}{%
    \textbf{Left:} Qualitative comparison on \emph{Rubble} (top) and \emph{Residence} (bottom): ground truth, \methodname{}, and a representative large-scale baseline. The PSNR annotated on each panel is for the single displayed view; scene-average PSNR is reported in Table~\ref{tab:main_results}.
    \textbf{Right:} Training time vs.\ PSNR on \emph{Residence} ($4\times$A6000 GPUs). \methodname{} trains much faster while maintaining high quality.}%
  \label{fig:teaser}%
  \vskip 1.0em
}
\makeatletter
\let\aaai@origmaketitlebody\@maketitle
\renewcommand{\@maketitle}{\aaai@origmaketitlebody\teaserfig}
\makeatother

\title{\methodname{}: City-Scale Gaussian Splatting at Lightning Speed}

\author{
    \normalfont
    Zhongtao Wang\textsuperscript{\textdagger}\qquad
    Huishan Au\textsuperscript{\textdagger}\qquad
    Yilong Li\qquad Mai Su\qquad Haojie Jin\\
    Yisong Chen\textsuperscript{*}\qquad
    Meng Gai\qquad
    Fei Zhu\textsuperscript{*}\qquad
    Guoping Wang
}
\affiliations{}

\begin{document}

\nocopyright
\maketitle
\begingroup
\renewcommand{\thefootnote}{\fnsymbol{footnote}}
\footnotetext[2]{Zhongtao Wang and Huishan Au contributed equally.}
\footnotetext[1]{Yisong Chen and Fei Zhu are the corresponding authors.}
\endgroup

\begin{abstract}
3D Gaussian Splatting underpins digital twins, simulation, and aerial mapping, yet city-scale training stays expensive even on multiple GPUs: every iteration must preprocess, communicate, and rasterize an overly dense set of primitives. At each step only a small fraction contributes to the loss, and the rest add pure storage, communication, and rasterization overhead. Existing methods target individual costs but not the underlying question: \emph{which Gaussians to store on each GPU, render for each view, and keep after early geometry forms?}

We present \textbf{\methodname{}}, a distributed 3DGS framework that reduces the active Gaussian workload at three coupled levels. At the \emph{system level}, it shards Gaussians across GPUs by index parity rather than spatial blocks, avoiding cross-block redundancy, and distributes each render through one cross-GPU exchange to the tile owners. At the \emph{model level}, it controls the population from both ends of densification: importance-scoring passes prune redundant Gaussians, a lightweight spawn gate withholds candidates predicted not to survive, and the same signal reweights density control. At the \emph{view level}, a distance-based LOD gate and an importance-based mask trim each camera's active set. On large-scale benchmarks, \methodname{} matches recent baselines in quality while training nearly an order of magnitude faster, in tens of minutes.
\end{abstract}

\section{Introduction}
\label{sec:intro}

City-scale 3D reconstruction has become a foundation for applications such as aerial survey inspection, digital twins, and simulation assets. These applications often begin with thousands of calibrated images, and end up with high-quality free-viewpoint novel views. A city-scale reconstruction system needs to build a scene quickly, use multiple GPUs efficiently, and render robustly across a wide range of viewing distances. While 3DGS works well on small scenes, it scales poorly. The vanilla pipeline becomes too slow to be practical, and the resulting model often struggles to preserve distant structures and large spatial extents.

Recent large-scale 3DGS work has addressed this cost from several directions, including distributed training~\citep{gao2025citygsx, chen2024dogs}, hierarchical scene partitioning with LOD rendering~\citep{kerbl2024hierarchical, ren2024octree, liu2024citygaussian}, and partition-and-merge scaling~\citep{lin2024vastgaussian, liu2024citygaussian, su2025hug}. Even with these advances, city-scale optimization remains expensive, typically spanning several hours across multi-GPU nodes. Training time remains a major bottleneck, and existing methods still trade off scene quality for training speed.

The reason is that these methods each accelerate a different part of the pipeline but share a common inefficiency. In any single iteration of city-scale 3DGS training, only a small fraction of the Gaussians contribute to the rendered loss, while the rest still incur the full cost of training. Some sit on a different GPU from the tile being rendered and must be communicated across devices before they can contribute. Others are too fine for the current camera and pass through projection and rasterization without affecting any pixel. The rest are redundant clones from unnecessary densification that never actually contribute. The dominant cost of city-scale 3DGS is therefore not only the size of the Gaussian set, but the \emph{wasted work} on Gaussians that the current step should not touch or create.

Our observation is that fast city-scale reconstruction should minimise the number of Gaussians that enter the rendering pipeline. Globally, the model should remove redundant clone-and-split leftovers after densification so they do not survive into the rest of the training. Locally, each view should activate only the scales that are useful for its camera distance, so distant or sub-pixel Gaussians do not incur unnecessary rasterization cost. Doing this at city scale needs a carefully designed importance signal, because aerial scenes contain a few large background surfaces alongside the fine facade and roof detail that actually matters. The signal must therefore reward how much a Gaussian contributes \emph{per pixel} rather than how many pixels it covers, so the background cannot crowd out that detail.

We present \textbf{\methodname{}}, a distributed 3DGS framework that reduces the active Gaussian workload at three coupled levels. At the \emph{system level}, it shards Gaussians across GPUs by index parity rather than by spatial blocks, so each device holds a balanced scene-wide slice free of cross-block visibility waste, and distributes each render step through a single cross-GPU exchange. At the \emph{model level}, \methodname{} controls the population from both ends of densification, vetoing candidates predicted not to survive before they spawn and pruning the redundant Gaussians that remain, while the same scoring pass emits importance signals that steer density control and view-level filtering at no extra cost. At the \emph{view level}, a distance-based LOD gate drops Gaussians too fine for the current camera and an importance-based cull skips those with negligible cross-view contribution, together trimming each camera's active set. The three levels each reduce a different component of the overall cost, namely per-GPU load, global model size, and per-view active set, so the same iteration touches fewer primitives at every stage of training.

Across Mill-19, UrbanScene3D, and \emph{MatrixCity}, \methodname{} matches the rendering quality of recent large-scale baselines while training city-scale scenes in tens of minutes on four A6000 GPUs. As shown in Figure~\ref{fig:teaser}, this yields nearly an order-of-magnitude training speedup.

In summary, our contributions are threefold:
\begin{itemize}[leftmargin=1.4em, itemsep=2pt, topsep=2pt]
  \item \textbf{Insight.} A single principle drives fast city-scale 3DGS: each operation should process as few Gaussians as possible without degrading reconstruction quality.
  \item \textbf{Method.} A distributed city-scale pipeline that unifies sharding, importance-based simplification, and LOD filtering, and controls the Gaussian population from both ends of densification by pruning redundant Gaussians and vetoing candidates predicted not to survive before they spawn.
  \item \textbf{Experiment.} A near-order-of-magnitude training speedup over recent city-scale baselines at comparable quality, with ablations confirming each level's contribution.
\end{itemize}

\section{Related Work}
\label{sec:related}

\begin{figure*}[t]
  \centering
  \includegraphics[width=0.98\textwidth]{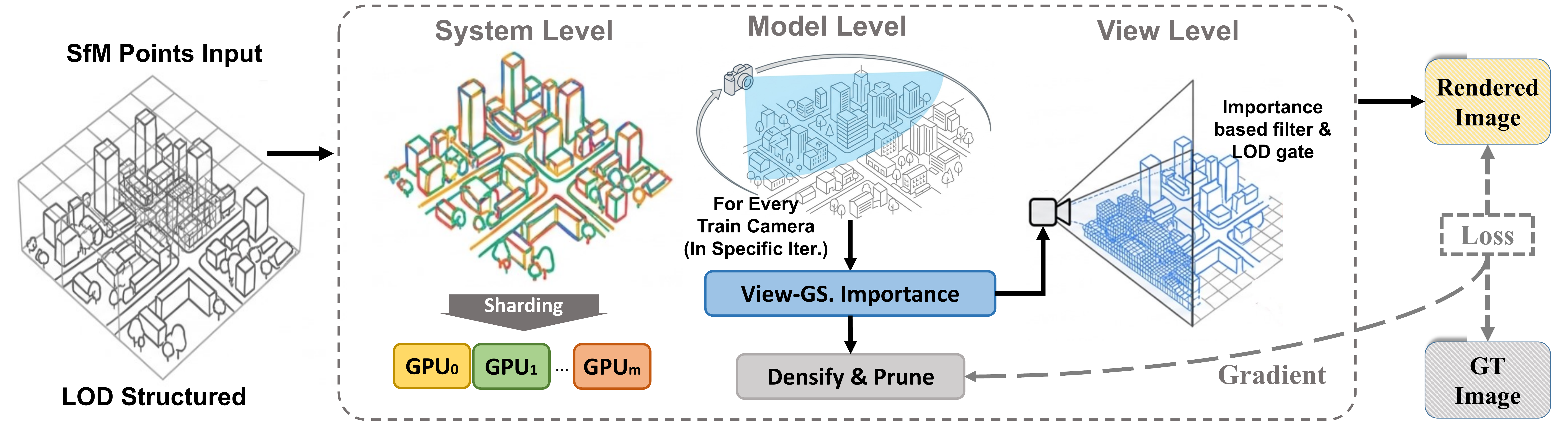}
  \caption{\textbf{Pipeline of \methodname{}.} The SfM cloud is voxelised at multiple scales, giving each Gaussian an LOD level. Three levels of active workload control then cut per-iteration cost at different scopes. The \emph{system level} shards Gaussians across $M$ GPUs non-spatially, so each rank holds a scene-wide slice, and distributes rendering across all ranks. The \emph{model level} interleaves scheduled simplification with density control. Each importance-scoring pass shrinks the population and emits a per-(Gaussian,~camera) signal that reweights density-control gradients and feeds the view level filter, while a spawn gate withholds vetoed candidates, limiting the peak population. The \emph{view level} trims each camera's active set with a distance-based LOD gate and an importance-driven cull. Training uses an L1+SSIM photometric loss with a scale regulariser on visible Gaussians.}
  \label{fig:method_pipeline}
\end{figure*}

\begin{figure}[t]
  \centering
  \includegraphics[width=0.98\columnwidth]{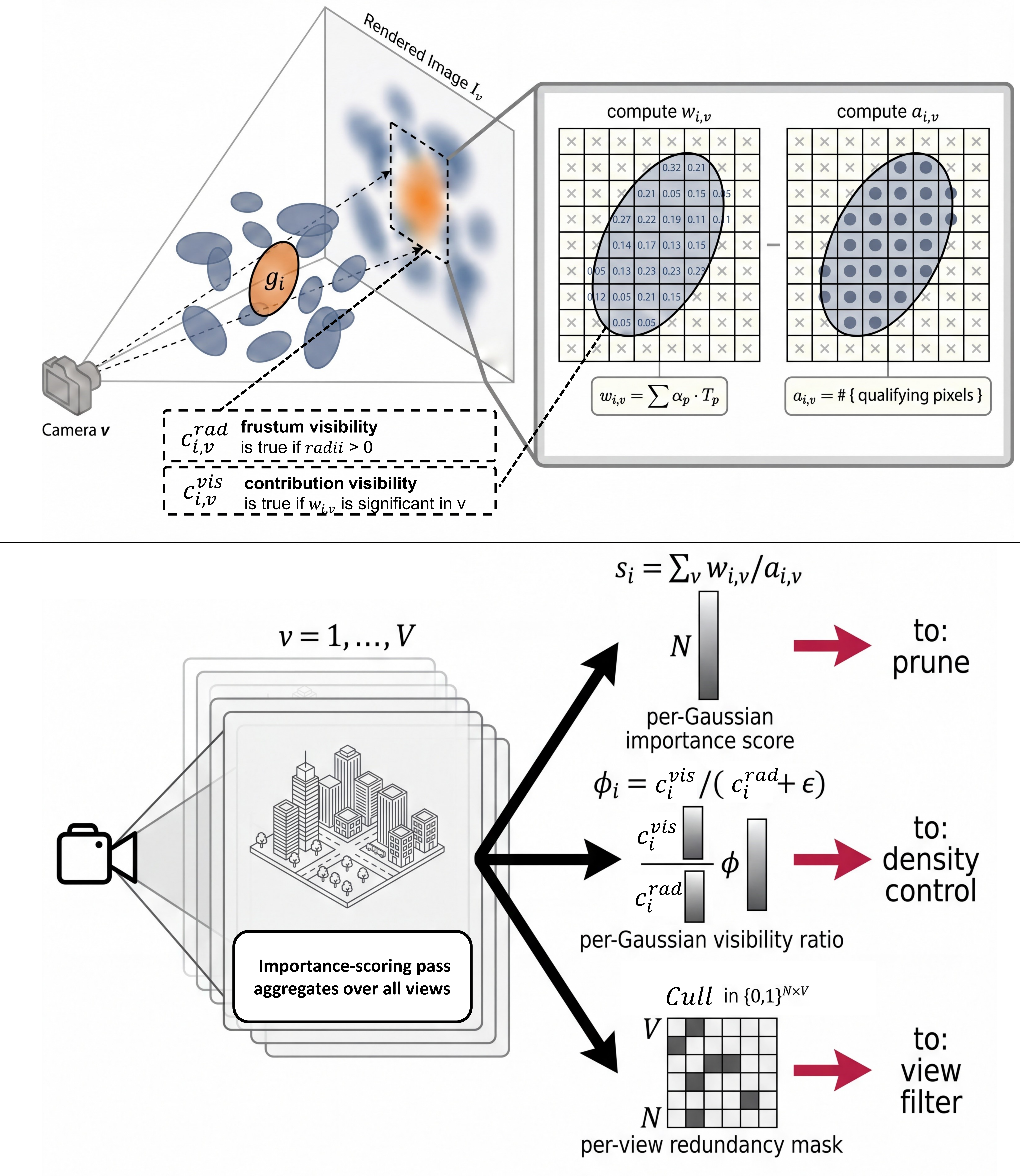}
  \caption{\textbf{Inside one importance-scoring pass.} One instrumented sweep over all $V$ views produces three signals at once. Per camera $v$ (top), the rasterizer records for each Gaussian $g_i$ the alpha-weight $w_{i,v}=\sum_p \alpha_p T_p$ and the qualifying-pixel count $a_{i,v}$. It also sets two visibility bits: $c^{\mathrm{rad}}_{i,v}$ (in frustum, $\text{radii}>0$) and $c^{\mathrm{vis}}_{i,v}$ ($w_{i,v}$ in the view's top-$99\%$ mass). Views are then aggregated (bottom). The score $s_i = \sum_{v:\,a_{i,v}>0} w_{i,v}/(a_{i,v}+\varepsilon)$ drives pruning. The ratio $\phi_i = c^{\mathrm{vis}}_i/(c^{\mathrm{rad}}_i + \varepsilon)$ reweights density control. The mask $\mathit{Cull}\in\{0,1\}^{N\times V}$ (complement of $c^{\mathrm{vis}}$) feeds the view level filter.}
  \label{fig:simplification}
\end{figure}

\begin{figure}[t]
  \centering
  \includegraphics[width=0.98\columnwidth]{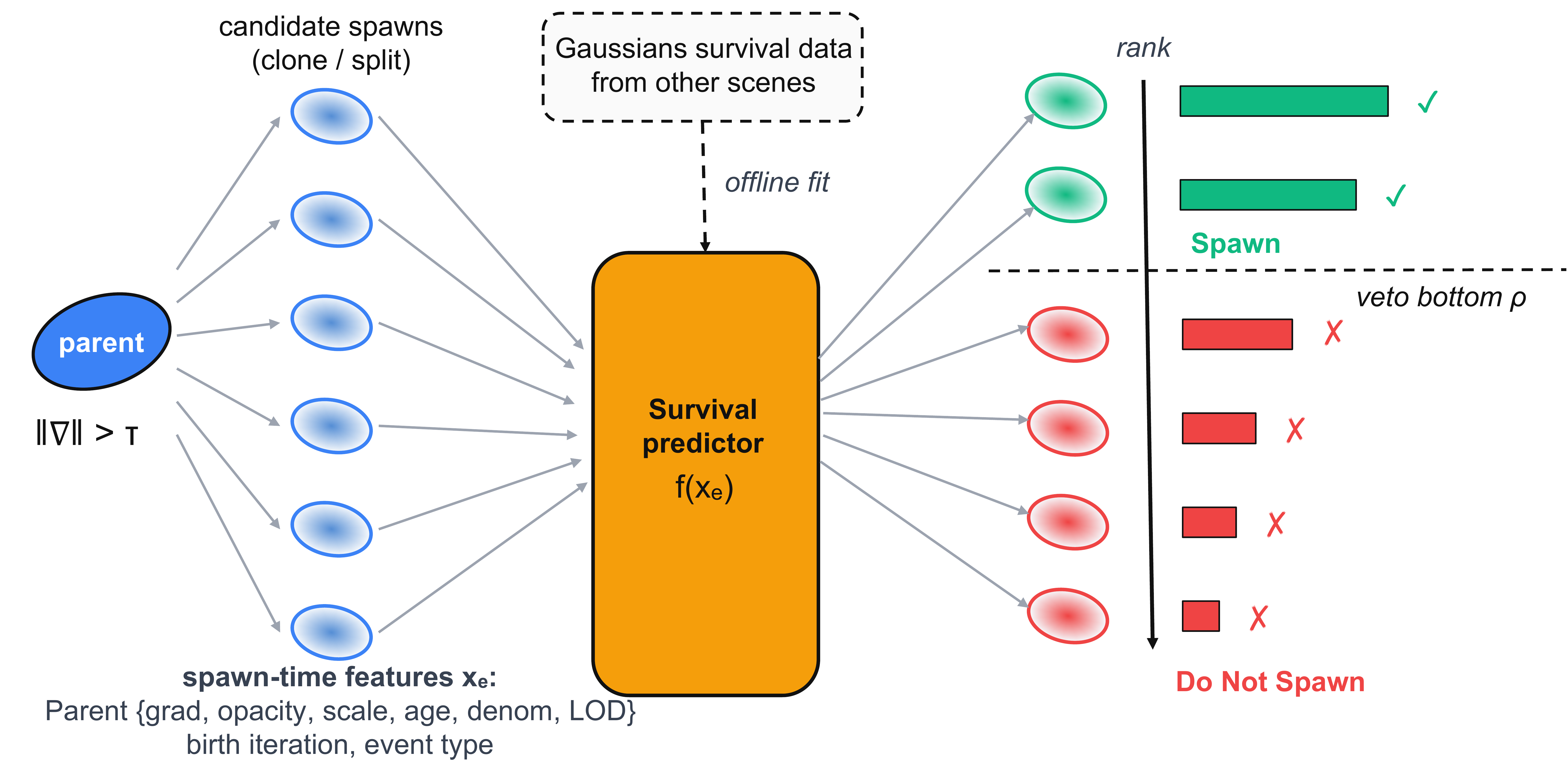}
  \caption{\textbf{The spawn gate.} At spawn time, a survival predictor fit offline on \emph{other} scenes scores each clone/split candidate. The bottom fraction $\rho$ is vetoed and never instantiated.}
  \Description{Left-to-right schematic of one gated densification event. On the left, parent Gaussians pass the gradient-threshold test and propose a column of candidate spawns. In the middle, each candidate's spawn-time feature vector enters a survival predictor box; a dashed arrow from above marks that the predictor is fit offline on lineage logs of other scenes (leave-one-scene-out). On the right, the candidates are re-ordered by predicted survival probability shown as horizontal score bars; the top candidates carry green check marks and an arrow labelled instantiate into the model, while the bottom fraction rho carries red crosses and the label vetoed, never instantiated.}
  \label{fig:natal_gate}
\end{figure}

\subsection{3D Gaussian Splatting}
3D Gaussian Splatting (3DGS)~\citep{kerbl3Dgaussians} represents scenes with explicit anisotropic primitives and differentiable tile-based rasterization, matching neural radiance fields~\citep{mildenhall2020nerf, muller2022instant, barron2022mipnerf360, fridovich2022plenoxels} in quality at real-time rates, yet its training still relies on progressive densification over tens of thousands of iterations. Follow-up work improves the representation along complementary axes: structured anchor or LOD variants~\citep{lu2024scaffold, ren2024octree}, anti-aliasing~\citep{yu2024mip}, densification and optimization~\citep{ye2024absgs, zhang2024pixelgs, cheng2024gaussianpro, kheradmand20243d, bulo2024revising}, geometric fidelity~\citep{huang20242dgs, chen2024pgsr, guedon2023sugar, Yu2024GOF}, and compactness via compression or pruning~\citep{lee2024compact, fan2023lightgaussian, liu2024maskgaussian}. These approaches target quality, geometry, or storage; \methodname{} instead targets the training workload at city scale.

\subsection{Large-Scale Scene Reconstruction}

Scaling to city-scale captures is commonly handled by decomposition. NeRF systems partition scenes or route samples to local models~\citep{tancik2022block, turki2022mega, zhu2023switchnerf}, and 3DGS pipelines follow suit: VastGaussian~\citep{lin2024vastgaussian} uses progressive data partitioning, CityGaussianV1~\citep{liu2024citygaussian} pairs block-wise reconstruction with LOD rendering, CityGaussianV2~\citep{liu2024citygaussianv2} improves geometry and compression, HUG~\citep{su2025hug} partitions by visibility with hierarchical urban Gaussians, and GigaGS~\citep{chen2024gigags} and LetsGo~\citep{cui2024letsgo} extend to high-fidelity surface and LiDAR-assisted capture. Such block pipelines must manage overlap, boundary consistency, and a final merge. Distributed and hierarchical methods reduce this more directly: RetinaGS~\citep{li2024retinags} trains billion-scale models model-parallel, Hierarchical 3DGS~\citep{kerbl2024hierarchical} builds a multi-scale hierarchy, DOGS~\citep{chen2024dogs} uses ADMM-style consensus, Momentum-GS~\citep{fan2025momentumgs} enforces block consistency via momentum self-distillation, and CityGS-X~\citep{gao2025citygsx} replaces partition-and-merge with a parallel hybrid hierarchical representation. By contrast, \methodname{} reduces the Gaussian workload per device, per view, and throughout training to reach minute-level reconstruction.

\subsection{Efficient Gaussian Splatting Optimization}

Accelerating 3DGS optimization is an active direction. Mini-Splatting~\citep{fang2024mini} and Mini-Splatting2~\citep{fang2025minisplatting2} speed up small-scene training through densification and simplification, Taming 3DGS~\citep{taming3dgs} steers pruning by importance, InstantSplat~\citep{fan2024instantsplat} uses feed-forward initialisation, and Grendel-GS~\citep{zhao2024scaling} adds distributed support for larger workloads via large-batch optimization~\citep{yao2018large}; a parallel line accelerates standard scenes through primitive pruning and resolution scheduling~\citep{niemeyer2025radsplat, hanson2025speedy, hanson2025pup, chen2025dashgaussian}. Closest to us, FastGS~\citep{ren2025fastgs} redesigns density control around multi-view-consistent densification and pruning, reaching around 100-second small-scene training; its strict Gaussian-count control mirrors our workload view, which \methodname{} extends to distributed city scale. Complementary systems work accelerates the renderer itself through rasterization memory reuse~\citep{feng2024flashgs}, faster gradient-pass atomics~\citep{durvasula2023distwar}, and popping-free sorting~\citep{radl2024stopthepop}. These approaches target small scenes or kernels; \methodname{} makes workload reduction the primary acceleration mechanism at city scale via much more efficient pruning. LeGS~\citep{ning2026legs} learns density control with per-scene reinforcement learning; our gate is supervised on lifetime outcomes and transfers across scenes.

\section{Method}
\label{sec:method}

Given posed aerial images $\{I_v\}_{v=1}^{V}$ and an SfM point cloud, \methodname{} reconstructs an explicit 3DGS model under distributed training. We keep the standard 3DGS parameterization throughout. A scene is a set of anisotropic primitives $\mathcal{G}=\{g_i\}_{i=1}^{N}$, each storing a center $\boldsymbol{\mu}_i$, a covariance $\boldsymbol{\Sigma}_i$, an opacity $\alpha_i$, and an SH color, rendered by tile-based front-to-back alpha compositing and optimized jointly with adaptive density control. We recall the formulation and notation in Appendix~\ref{sec:preliminaries}. What changes at city scale is where the cost lies. The dominant expense is no longer the number of images but the number of \emph{active} Gaussians that every iteration must project, exchange across devices, rasterize, and update. The method therefore follows a single rule: \textbf{a city-scale optimization should avoid touching Gaussians that are unnecessary for the current device, view, or training stage}. As shown in Figure~\ref{fig:method_pipeline}, we instantiate this rule at three coupled levels. The system level applies sharding with distributed rendering, the model level applies simplification that feeds an importance signal back to density control and view filtering, and the view level trims by LOD and importance.

\subsection{System Level Primitive Sharding}
\label{sec:system}

A city-scale model holds hundreds of millions of primitives, far more than a single GPU can store and optimize. Partitioning the scene into world-space blocks incurs cross-block visibility waste, imbalanced shards, and a final merge (Appendix~\ref{sec:sharding_appendix}). Inspired by \cite{zhao2024scaling,gao2025citygsx}, \methodname{} shards the global model across $M$ GPUs by index parity and distributes every render step across the same GPUs, so both storage and per-step compute scale with the GPU count.

Each GPU holds only its shard $\mathcal{G}^{(m)}$ ($\approx|\mathcal{G}|/M$ Gaussians) and optimizer state, and projects only its own Gaussians. The rendered image is tiled, with each tile assigned to one GPU. A single all-to-all step then routes every projected Gaussian to the GPUs owning the tiles it covers, which composite their pixels from what they receive. Projection and rasterization are thus distributed every step, while the output stays identical to a single-GPU renderer on the same set. Two load-balancers keep this productive: an index-parity redistribution rebalances shard sizes whenever the per-GPU skew exceeds a threshold, and a cost-aware tile partition evens out rasterization load.

\subsection{Model Level Simplification}
\label{sec:model}

System level sharding balances per-iteration work but does not reduce the global count $|\mathcal{G}|$. Densification drives $|\mathcal{G}|$ well above what the final scene needs, which significantly affects the training speed regardless of how well the workload is sharded. \methodname{} therefore controls the population at the model level from both ends of densification. \emph{Scheduled simplification passes} score each Gaussian by its contribution density (the importance score $s_i$ below) and permanently prune the redundant ones (Figure~\ref{fig:simplification}). At the other end, a lightweight \emph{spawn gate} vetoes candidates predicted at birth not to survive.

\paragraph{Importance score.} An importance-scoring pass is a single sweep over all $V$ training views with an instrumented rasterizer that bypasses color shading. For each view $v$ and each Gaussian $g_i$ projected onto pixels in $v$, the rasterizer records the alpha-weight $w_{i,v}$ that $g_i$ contributes summed across the pixels it touches, and the projected support area $a_{i,v}$, the number of pixels it touches. The pass aggregates these into a single per-Gaussian score $s_i = \sum_{v\,:\,a_{i,v}>0} w_{i,v}/(a_{i,v}+\varepsilon)$, where $\varepsilon$ is a small constant for numerical stability. Dividing by the area makes $s_i$ a per-pixel contribution, rather than a raw pixel count, which matters at city scale. Without it, a few large background Gaussians would score highly just by covering many pixels, and the small Gaussians that carry fine detail would be pruned instead. The score is recomputed from scratch in each pass, but never updated during regular training iterations.

\paragraph{Scheduled simplification and feedback.} We run two passes with complementary pruning rules. The first pass (early, at $T_1$) reduces the population by \emph{stochastic} importance-weighted sampling without replacement, drawing a fixed fraction with probability proportional to $s_i$. High-contribution primitives are retained almost every time, while low-scoring primitives are kept only occasionally, primarily safeguarding quality. The second pass (late, at $T_2$) applies a \emph{deterministic} cumulative-mass cut that keeps the smallest prefix whose summed score reaches $99\%$ of the total, primarily trimming the final population. Each pass is followed by an index-parity redistribution that rebalances survivors across shards. At no extra rendering cost, the same sweep also records two per-(Gaussian,~view) visibility bits---frustum visibility $c^{\text{rad}}_{i,v}$ and contribution visibility $c^{\text{vis}}_{i,v}$ ($w_{i,v}$ lies in the per-view top-$99\%$ mass), with view sums $c_i^{\text{rad}}=\sum_v c^{\text{rad}}_{i,v}$ and $c_i^{\text{vis}}=\sum_v c^{\text{vis}}_{i,v}$. The ratio $\phi_i = c_i^{\text{vis}}/(c_i^{\text{rad}}+\varepsilon)$ measures how often $g_i$ genuinely contributes when visible. Density control reweights the clone-and-split gradient by $\phi_i$, concentrating the remaining densification on consistently-contributing primitives. The complementary mask $\mathit{Cull}_{i,v}=1-c^{\text{vis}}_{i,v}$ goes to the view level filter (Sec.~\ref{sec:view}), flagging Gaussians outside each camera's top-$99\%$ mass. One sweep therefore drives \emph{all three roles at once}: shrinking the population, steering densification, and selecting each camera's active subset.

\paragraph{Spawn gate.}\label{sec:natal}
Every mechanism so far is \emph{reactive}: a Gaussian is born, optimized, communicated, and rasterized, and only then simplified away if it proves redundant. This reactive removal leaves the mid-training population \emph{peak} untouched, and that peak sets the per-GPU memory bottleneck. We find that in our method, the peak is inflated by late-born Gaussians that rarely survive, and their survival is already predictable at spawn (See Appendix~\ref{sec:lineage_appendix}). The \emph{spawn gate} exploits this to attack the peak \emph{proactively}. Rather than grow-then-prune, it predicts at birth which densification candidates will not survive and refuses to instantiate them. Their optimization, communication, and rasterization costs are then never paid. The gate filters candidates at each densification event. After the standard gradient test selects clone and split candidates, a gradient-boosted tree $f(\mathbf{x}_e)$ predicts each candidate's survival from features available at spawn time. This predictor is trained offline on exact per-Gaussian lifetime labels. The bottom fraction by rank ($\rho{=}0.6$) is then vetoed (Fig.~\ref{fig:natal_gate}). The gate activates only in the late densification window, where survival is lowest. As we analyse in Appendix~\ref{sec:lineage_appendix}, the gate is a natural fit for our aggressive schedule. Such schedules overproduce short-lived spawns, exactly what the gate removes, so it reclaims substantial peak memory.   Every scene in Sec.~\ref{sec:experiments} uses a predictor fit exclusively on \emph{other} scenes, so no gate sees its own data. Appendices~\ref{sec:lineage_appendix} and~\ref{sec:oracle_appendix} report the full supporting analysis, including the lifetime measurements, the leave-one-scene-out transfer, the rate and onset sensitivity, and the oracle-ceiling comparison, where the learned gate captures $98\%$ of a perfect-foresight oracle's peak-population cut with no measurable quality gap.

\subsection{View Level Filtering}
\label{sec:view}

The system level sharding distributes primitives across devices and the model level simplification trims the global set. At the view level, \methodname{} further trims the active set entering each render through two complementary filters: a \emph{distance-based LOD gate} and an \emph{importance-based per-view culling mask}.

\paragraph{Distance-based LOD gate.} Each Gaussian carries an LOD label $\ell_i \in \{0, \ldots, K-1\}$ attached at multi-resolution voxelisation of the SfM cloud~\citep{ren2024octree}. The label is propagated through density control by a simple inheritance rule: clone keeps the parent's level, split increments by one. The level structure therefore remains a stable property of the model rather than a render-time-only mask. For a camera with centre $\mathbf{c}_v$ at training iteration $t$, the renderer estimates the finest level resolvable at distance $d$,
\begin{equation}
  L_v(d,t)=\mathrm{clamp}\!\Bigl(\bigl\lfloor \log_f(d_0/d) \bigr\rceil,\, 0,\, L_{\max}(t)\Bigr),
  \label{eq:distance_level}
\end{equation}
and keeps only Gaussians with equal or lower levels:
\begin{equation}
  \mathcal{L}^{(m)}_v(t)=\{\,g_i\in\mathcal{G}^{(m)}\,:\,\ell_i\le L_v(\|\boldsymbol{\mu}_i-\mathbf{c}_v\|,t)\,\}.
  \label{eq:lod_active_set}
\end{equation}
$d_0$ is a reference camera-to-scene distance computed from the training cameras at initialisation and held fixed, and $f=2$ is the geometric ratio between consecutive LOD levels. $L_{\max}(t)$ unlocks levels coarse-to-fine on a geometric schedule so the LOD set's ceiling tracks the population that has actually formed. The gate itself is disabled throughout the density-control window, so masked-out Gaussians do not lose densification credit, and engages once density control ends. We fall back to the full shard whenever the predicted ratio $|\mathcal{L}^{(m)}_v(t)|/|\mathcal{G}^{(m)}|$ is close to one, so as to skip the gate when it would barely filter anything.

\paragraph{Importance-based per-view culling mask.} The importance-scoring passes of Sec.~\ref{sec:model} maintain, alongside the per-Gaussian score, a bit-packed per-view redundancy matrix $\mathit{Cull}\in\{0,1\}^{N\times V}$ that flags, for every camera, the Gaussians whose contribution to that camera fell below the per-view top quantile in the most recent sweep. Its rows are sharded with the Gaussians and bit-packed, so its footprint is negligible. Whenever this mask is available, the main render consults the column corresponding to the current camera and skips the flagged Gaussians on top of the LOD gate, so the per-rank active set actually rasterized for view $v$ becomes
\begin{equation}
  \mathcal{A}^{(m)}_v(t)=\mathcal{L}^{(m)}_v(t)\setminus\{\,g_i:\mathit{Cull}_{i,v}=1\,\}.
  \label{eq:active_set}
\end{equation}
This filter and the LOD gate together apply two independent reductions to the per-camera active set: a geometric one driven by camera distance, and a contribution-based one driven by cross-view importance.

\subsection{Supervision}
\label{sec:supervision}

For a mini-batch of $B$ training cameras, each rank supervises only the image tiles it owns. The tile losses are reduced across ranks, and gradients propagate through both the rasterizer and the screen-space routing of Sec.~\ref{sec:system}. The photometric term is the standard RGB reconstruction loss
\begin{equation}
  \mathcal{L}_{\mathrm{photo}}=\frac{1}{B}\sum_{b=1}^{B}
  (1-\lambda)\|\hat{I}_b-I_b\|_1+
  \lambda\bigl(1-\mathrm{SSIM}(\hat{I}_b,I_b)\bigr),
  \label{eq:loss_rgb}
\end{equation}
where $\lambda \in [0,1]$ blends the L1 and SSIM terms. Following common practice in surface-aware 3DGS methods~\citep{chen2024pgsr}, we add a regulariser that penalises the smallest scale axis of each Gaussian in the visible set $\mathcal{V}$ to encourage surface-aligned primitives:
\begin{equation}
  \mathcal{L}=\mathcal{L}_{\mathrm{photo}}+\beta\,\mathcal{L}_{\mathrm{scale}},\qquad
  \mathcal{L}_{\mathrm{scale}}=\frac{1}{|\mathcal{V}|}\sum_{i\in\mathcal{V}}\min_{j}s_{i,j}.
  \label{eq:loss_total}
\end{equation}
Here $\beta$ weighs the scale regulariser and $s_{i,j}$ is the scale of $g_i$ along its $j$-th principal axis of the covariance $\boldsymbol{\Sigma}_i$. The acceleration in \methodname{} comes from controlling the active Gaussian workload, not from auxiliary supervision. We therefore keep this objective lightweight and apply it consistently across all distributed tiles.

\section{Experiments}
\label{sec:experiments}

\begin{table*}[t]
\centering

\scriptsize
\setlength{\tabcolsep}{2.2pt}
\resizebox{\textwidth}{!}{%
\begin{tabular}{l|cccc|cccc|cccc|cccc|cccc}
\toprule
\multirow{2}{*}{Method} &
  \multicolumn{4}{c|}{\emph{Building}} &
  \multicolumn{4}{c|}{\emph{Rubble}} &
  \multicolumn{4}{c|}{\emph{Residence}} &
  \multicolumn{4}{c|}{\emph{Sci-Art}} &
  \multicolumn{4}{c}{\emph{MatrixCity}} \\
\cmidrule{2-21}
 & PSNR$\uparrow$ & SSIM$\uparrow$ & LPIPS$\downarrow$ & Time
 & PSNR$\uparrow$ & SSIM$\uparrow$ & LPIPS$\downarrow$ & Time
 & PSNR$\uparrow$ & SSIM$\uparrow$ & LPIPS$\downarrow$ & Time
 & PSNR$\uparrow$ & SSIM$\uparrow$ & LPIPS$\downarrow$ & Time
 & PSNR$\uparrow$ & SSIM$\uparrow$ & LPIPS$\downarrow$ & Time \\
\midrule
VastGaussian~\citep{lin2024vastgaussian}
  & 21.80 & 0.728 & 0.225 & 5h48m
  & 25.20 & 0.742 & 0.264 & \rkB{2h16m}
  & 21.01 & 0.699 & 0.261 & 5h05m
  & -- & -- & -- & --
  & -- & -- & -- & -- \\
CityGaussianV1~\citep{liu2024citygaussian}
  & 21.55 & 0.778 & 0.246 & 8h07m
  & 25.77 & 0.813 & 0.228 & 3h50m
  & \rkC{22.00} & \rkC{0.813} & 0.211 & 8h33m
  & 21.39 & 0.837 & 0.230 & 3h56m
  & \rkC{27.46} & \rkC{0.865} & 0.204 & 12h17m \\
CityGaussianV2~\citep{liu2024citygaussianv2}
  & 22.23 & 0.759 & 0.217 & 7h01m
  & 24.58 & 0.767 & 0.252 & 4h16m
  & 21.71 & 0.780 & 0.225 & 7h45m
  & 21.49 & 0.811 & 0.238 & 10h35m
  & 27.23 & 0.857 & \rkB{0.169} & 8h48m \\
Momentum-GS~\citep{fan2025momentumgs}
  & \rkB{23.19} & \rkB{0.816} & \rkB{0.193} & 7h04m
  & \rkC{25.91} & 0.829 & \rkB{0.197} & 5h03m
  & 21.18 & 0.760 & 0.248 & 9h09m
  & 20.75 & 0.794 & 0.266 & 6h27m
  & \rkA{28.07} & \rkB{0.879} & \rkC{0.183} & \rkC{6h14m} \\
HUG~\citep{su2025hug}
  & \rkC{22.35} & \rkC{0.792} & 0.228 & 5h16m
  & \rkB{26.42} & \rkB{0.839} & \rkB{0.197} & \rkC{2h37m}
  & \rkB{22.33} & \rkC{0.813} & \rkC{0.207} & 6h31m
  & 21.83 & 0.846 & 0.204 & \rkC{2h58m}
  & \rkB{28.02} & \rkA{0.883} & \rkA{0.142} & 10h04m \\
CityGS-X~\citep{gao2025citygsx}
  & 22.14 & \rkB{0.816} & \rkA{0.186} & \rkC{4h30m}
  & 24.92 & \rkC{0.831} & \rkC{0.199} & 6h03m
  & 21.49 & \rkB{0.828} & \rkA{0.177} & \rkC{4h45m}
  & \rkA{23.31} & \rkB{0.872} & \rkB{0.178} & 3h14m
  & 27.02 & 0.852 & 0.240 & 9h18m \\
Grendel-GS (paper config)~\citep{zhao2024scaling}
  & 19.27 & 0.673 & 0.326 & \rkB{2h13m}
  & 24.57 & 0.780 & 0.251 & 3h23m
  & 21.18 & 0.780 & 0.223 & \rkB{3h14m}
  & \rkC{22.68} & \rkC{0.856} & \rkC{0.188} & \rkB{2h16m}
  & 23.82 & 0.711 & 0.420 & \rkB{1h38m} \\
Grendel-GS (long iter)~\citep{zhao2024scaling}
  & 19.50 & 0.678 & 0.319 & 6h05m
  & 25.26 & 0.792 & 0.242 & 8h06m
  & 21.32 & 0.784 & 0.219 & 8h10m
  & 22.66 & 0.851 & 0.197 & 5h38m
  & 23.20 & 0.693 & 0.427 & 8h18m \\
\midrule
\textbf{\methodname{} (Ours)}
  & \rkA{23.46} & \rkA{0.819} & \rkC{0.207} & \rkA{\textbf{39m15s}}
  & \rkA{27.91} & \rkA{0.859} & \rkA{0.194} & \rkA{\textbf{37m36s}}
  & \rkA{22.67} & \rkA{0.844} & \rkB{0.184} & \rkA{\textbf{37m47s}}
  & \rkB{22.92} & \rkA{0.882} & \rkA{0.169} & \rkA{\textbf{34m33s}}
  & 26.87 & 0.845 & 0.256 & \rkA{\textbf{1h13m}} \\
\bottomrule
\end{tabular}%
}
  \caption{\textbf{Quality and reconstruction time on large-scale benchmarks.}
We compare \methodname{} with baselines on standard novel-view synthesis metrics and training time. Per metric and scene, the top three methods are highlighted with \colorbox{red!30}{red}\,/\,\colorbox{orange!30}{orange}\,/\,\colorbox{yellow!30}{yellow} cell backgrounds for 1st\,/\,2nd\,/\,3rd respectively. Dashes (--) mark cells we could not reproduce because the corresponding baseline did not release a public checkpoint or training code for that scene. Grendel-GS is shown at two settings (Appendix~\ref{sec:impl_appendix}).}
  \label{tab:main_results}
\end{table*}

\begin{table}[t]
\centering

\footnotesize
\setlength{\tabcolsep}{3pt}
\resizebox{\columnwidth}{!}{%
\begin{tabular}{l|c|ccc}
\toprule
Variant & PSNR$\uparrow$ & it/s$\uparrow$ & Time$\downarrow$ & \#GS$_{\text{pk}\to\text{fn}}$\,(M)$\downarrow$ \\
\midrule
Full \methodname{}                          & 23.46 & 36.8 & 39m15s & 8.66$\to$4.65 \\
\;w/o octree LOD                             & 23.57 & 31.6 & 45m32s & 10.86$\to$5.96 \\
\;w/o importance scoring                     & 23.30 & 22.2 & 64m05s & 10.93$\to$10.93 \\
\;w/o per-view importance mask               & 23.32 & 34.7 & 41m44s & 8.66$\to$4.64 \\
\;w/o $\phi_i$ density-control reweighting   & 23.21 & 35.9 & 40m10s & 9.55$\to$4.98 \\
\;w/o Pass-1 pruning                         & 23.30 & 34.9 & 41m25s & 10.01$\to$4.80 \\
\;w/o Pass-2 pruning                         & 23.43 & 27.7 & 51m16s & 8.68$\to$8.67 \\
\;w/o spawn gate                             & 23.40 & 35.4 & 40m48s & 10.89$\to$4.96 \\
\;all off                                    & 23.22 & 19.7 & 71m46s & 13.49$\to$13.48 \\
\;partition ($4{\times}5$, 20 blocks)        & 21.43 & -- & 115m36s & --$\to$5.37 \\
\;partition ($5{\times}6$, 30 blocks)        & 22.43 & -- & 169m48s & --$\to$5.43 \\
\bottomrule
\end{tabular}%
}
  \caption{\textbf{Component analysis on \emph{Building} (Mill-19), at 90k iterations.}
Each row removes one component from full \methodname{}; \emph{all off} removes all of them. For the two block-partitioning rows, Time is equivalent wall-clock (split $+$ four-GPU makespan $+$ merge), and it/s and global peak are undefined for independent per-block training (--). \#GS is live Gaussians as peak$\to$final. The ``w/o Pass-X pruning'' variants skip only that pass's population-reduction step, keeping its signal computations ($s_i$, $\phi_i$, $\mathit{Cull}$) intact.}
  \label{tab:workload_ablation}
\end{table}

\subsection{Experimental Protocol}

\paragraph{Datasets.}
We evaluate on the standard aerial benchmarks of large-scale 3DGS reconstruction: \emph{Building} ($1{,}920$ train images) and \emph{Rubble} ($1{,}678$) from Mill-19~\citep{turki2022mega}, \emph{Residence} ($2{,}582$) and \emph{Sci-Art} ($3{,}019$) from UrbanScene3D~\citep{lin2022capturing}, and the aerial split of \emph{MatrixCity} ($5{,}621$)~\citep{li2023matrixcity}. We follow the train/test camera splits released by Mega-NeRF~\citep{turki2022mega} and UrbanScene3D, and the official aerial split of \emph{MatrixCity}. 

\paragraph{Baselines and metrics.}
We compare \methodname{} against representative large-scale 3DGS baselines: VastGaussian~\citep{lin2024vastgaussian}, CityGaussianV1~\citep{liu2024citygaussian}, CityGaussianV2~\citep{liu2024citygaussianv2}, HUG~\citep{su2025hug}, Momentum-GS~\citep{fan2025momentumgs}, CityGS-X~\citep{gao2025citygsx} and Grendel-GS~\citep{zhao2024scaling}. We report PSNR, SSIM, and LPIPS-VGG for novel-view synthesis on the held-out test cameras, together with training time. For baselines with public checkpoints~\cite{lin2024vastgaussian,liu2024citygaussian}, we evaluate quality from the official checkpoint and measure training time by rerunning their released code on our hardware. The reported time covers the full pipeline end to end, including all GPU communication overhead.

\paragraph{Implementation details.}
All results are produced on 4$\times$A6000 GPUs with batch size $B{=}4$ unless otherwise stated. \methodname{} trains for $90\mathrm{k}$ iterations ($120\mathrm{k}$ on \emph{MatrixCity}). The two importance-scoring passes run at $T_1{=}16\mathrm{k}$ and $T_2{=}34\mathrm{k}$, the spawn gate is active from iteration $24\mathrm{k}$ at veto rate $\rho{=}0.6$, and the loss weights are $\lambda{=}0.2$ (L1--SSIM) and $\beta{=}10$ (scale regulariser). Further schedule details and baseline configurations are in Appendix~\ref{sec:impl_appendix}.

\subsection{Quality-Speed Comparison}

\begin{figure*}[t]
  \centering
  \includegraphics[width=\textwidth]{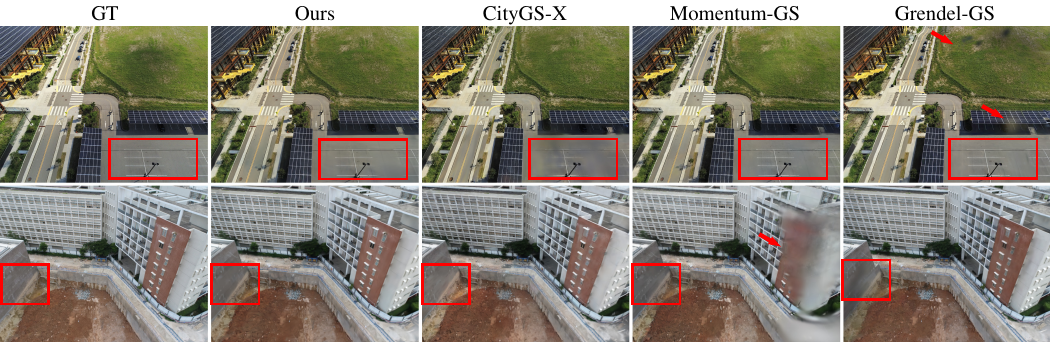}
  \caption{\textbf{Qualitative comparison on two held-out test views} from \emph{Building} (top) and \emph{Sci-Art} (bottom). Red rectangles mark the close-up regions, and red arrows highlight artifacts and blurred detail in the baselines. Additional comparisons are provided in Appendix~\ref{sec:qualitative_appendix}.}
  \Description{A two-row by five-column qualitative comparison. Columns are labelled, from left to right, GT, Ours, CityGS-X, Momentum-GS and Grendel-GS. The top row is an aerial Building view of a solar-panel roof running alongside a green lawn and a parking lot, with a red rectangle over a stretch of pavement holding a street lamp, its shadow and faint parking-line markings; BlitzGS keeps the lamp and lines crisp and matches GT, while CityGS-X blurs the boxed pavement and mottles it with uneven color, and Grendel-GS scatters dark floater blobs over the lawn and near the solar panels that the red arrows mark. The bottom row is an aerial Sci-Art view of a grey building on the left beside a white building with a red-brick wing, fronted by a bare reddish-brown earth ground, with a red rectangle over the lower-left corner where a retaining wall meets the earth; BlitzGS preserves the wall-and-debris boundary and stays closest to GT, while CityGS-X replaces the boxed corner with a washed-out smear, Momentum-GS blurs the red-brick wing into a smeared grey-white patch that the red arrow marks, and Grendel-GS blurs the grey retaining wall. The figure conveys that BlitzGS matches the strongest baselines on detail despite an order-of-magnitude shorter training time, and that CityGS-X, Momentum-GS and Grendel-GS each introduce visible artifacts on these views.}
  \label{fig:visual_two_rows}
\end{figure*}

Table~\ref{tab:main_results} shows that \methodname{} matches the rendering quality of recent large-scale 3DGS baselines across both real and synthetic aerial scenes while reducing training time by nearly an order of magnitude. \methodname{} achieves the best overall rendering quality on 4 real-world datasets, and stays competitive on the synthetic \emph{MatrixCity} scene. Table~\ref{tab:workload_ablation} additionally reports peak and final Gaussian counts.

\begin{figure}[t]
  \centering
  \includegraphics[width=0.99\columnwidth]{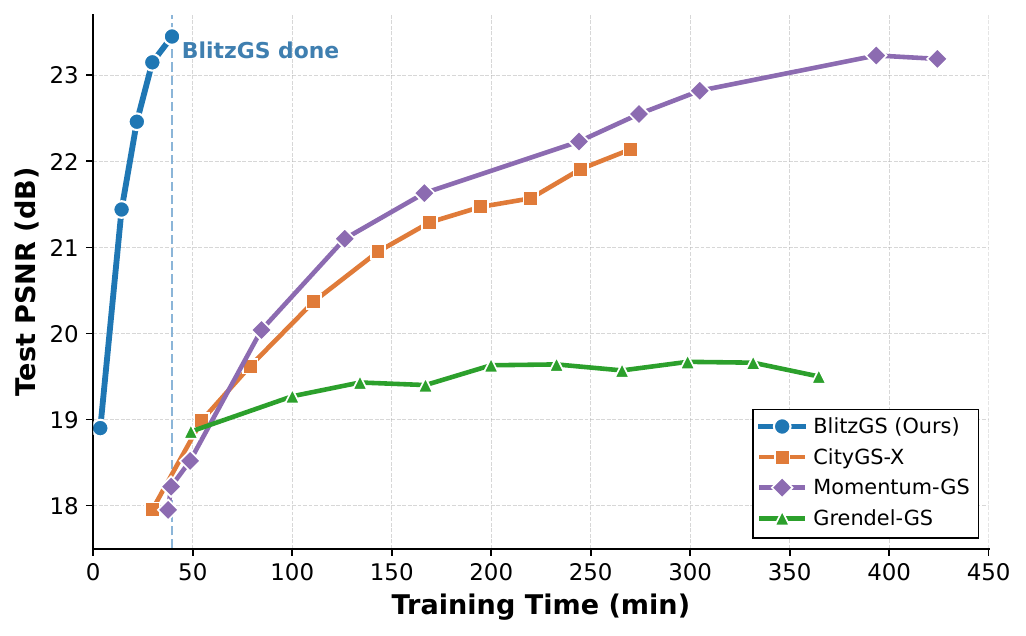}
  \caption{\textbf{Test PSNR over training time (\emph{Building}).} \methodname{} reaches a competitive PSNR in minutes (dashed line); the baselines train for hours, and Grendel-GS plateaus about $4$~dB lower.}
  \Description{A test-PSNR versus training-time convergence plot on the \emph{Building} scene comparing four large-scale 3DGS systems. The horizontal axis spans 0 to about 450 minutes; the vertical axis spans roughly 17.5 to 23.7 dB PSNR. BlitzGS is plotted as a short, steep blue curve with circle markers, climbing from about 18.9 dB at 4 minutes to about 23.5 dB at 39 minutes and terminating at a vertical light-blue dashed line labelled BlitzGS done. CityGS-X is drawn as an orange curve with square markers extending from about 30 to 270 minutes and rising gradually to roughly 22.1 dB. Momentum-GS appears as a purple curve with diamond markers stretching from about 38 to 424 minutes and ending near 23.2 dB. Grendel-GS is drawn as a green curve with triangle markers from about 49 to 365 minutes that stays flat near 19.5 dB, well below every other method. The visual conveys that BlitzGS reaches the eventual plateau of the baselines an order of magnitude earlier in wall-clock time, and that Grendel-GS never reaches a competitive quality despite the longest training.}
  \label{fig:convergence_building}
\end{figure}

\begin{figure}[t]
  \centering
  \includegraphics[width=0.98\columnwidth]{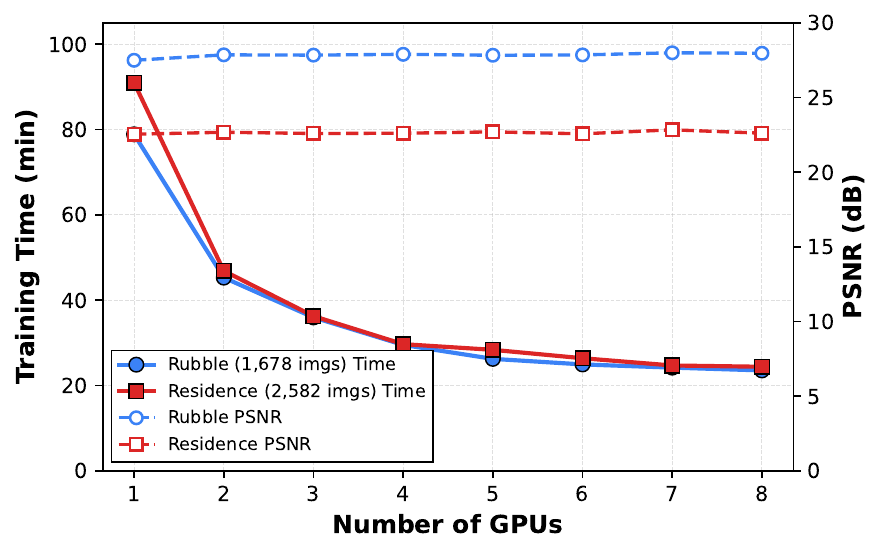}
  \caption{\textbf{Multi-GPU scaling across dataset sizes.} Color identifies the dataset (\emph{Rubble}, blue; \emph{Residence}, red); solid lines and filled markers are training time (left axis), dashed lines and hollow markers are PSNR (right axis). Time drops steadily while PSNR stays flat on both scenes.}
  \Description{A single-panel multi-GPU scaling plot with twin vertical axes, comparing BlitzGS on \emph{Rubble} (1,678 images, blue, circle markers) and \emph{Residence} (2,582 images, red, square markers) from one to eight A800 80\,GB GPUs (a single 48\,GB A6000 does not fit the undistributed $M{=}1$ configuration). The left axis shows training time in minutes plotted as solid lines with filled markers: both curves descend monotonically and almost in parallel, \emph{Rubble} dropping from 79 to 24 minutes and \emph{Residence} from 91 to 24 minutes, indicating consistent speedup across dataset sizes. The right axis shows test PSNR in dB plotted as dashed lines with hollow markers: each scene's curve is essentially flat, \emph{Rubble} varying within a narrow 27.5-28.0 dB band and \emph{Residence} within a 22.5-22.8 dB band, indicating that scaling does not degrade reconstruction quality. A two-column legend in the upper area identifies the four series. The plot conveys that BlitzGS scales effectively whether the input set contains roughly two thousand or two and a half thousand images.}
  \label{fig:scalability}
\end{figure}

To make the time-to-quality trade-off explicit, we additionally compare \methodname{} against CityGS-X, Momentum-GS, and Grendel-GS on \emph{Building}. We checkpoint each method at fixed iteration intervals and evaluate held-out test PSNR from every checkpoint once all runs finish (Figure~\ref{fig:convergence_building}). \methodname{} reaches about $23.5$~dB in roughly $39$ minutes, whereas CityGS-X and Momentum-GS keep training for $4$--$6$ more hours. Momentum-GS eventually catches up, while CityGS-X plateaus about $1.3$~dB lower. Grendel-GS trains for six hours yet plateaus near $19.5$~dB, confirming that scaling up distributed training without workload reduction does not reach city-scale quality. \methodname{} thus matches or surpasses their eventual plateau roughly an order of magnitude earlier in wall-clock time. Each iteration yields more PSNR per second, consistent with the workload model of Sec.~\ref{sec:method}.

\subsection{Qualitative Results}

As shown in Figure~\ref{fig:visual_two_rows}, \methodname{} preserves detail better than the baselines while training an order of magnitude faster. Additional comparisons are in Appendix~\ref{sec:qualitative_appendix} (Figures~\ref{fig:visual_comparison} and~\ref{fig:visual_at_35min}).

\subsection{Ablation Study}

We next examine how each level of workload control influences the speed-quality trade-off. Unless otherwise stated, all variants in this section use the default hyperparameters of Sec.~\ref{sec:method} and toggle only the marked component.

Table~\ref{tab:workload_ablation} reports the ablation. Removing the importance-scoring pass is the most damaging. With no pruning the population never shrinks and training takes $63\%$ longer. Removing the octree LOD inflates the final population by $28\%$. The per-view importance mask has only a marginal effect once the LOD gate is in place, since the two filters overlap on Gaussians outside the camera's resolvable range. Among the model-level mechanisms, removing the $\phi_i$ reweighting costs $0.25$~dB PSNR, confirming its role in concentrating densification onto view-contributory primitives. Removing Pass~1 pruning (the early stochastic sampling pass at $T_1$) costs $0.16$~dB. Removing Pass~2 pruning (the late deterministic $99\%$-mass cut at $T_2$) leaves PSNR nearly unchanged but nearly doubles the final population, marking it as the long-tail trim. The spawn gate is quality-neutral but trims the mid-training \emph{peak} population by about $20\%$, lowering per-GPU peak memory (Sec.~\ref{sec:natal}). Turning off every component at once (\emph{all off}) is far slower at similar quality: together the components make training about $1.8\times$ faster and the model $3\times$ smaller at comparable quality. The last two rows of Table~\ref{tab:workload_ablation} replace index-parity sharding with spatial block partitioning, all else fixed. Quality stays below \methodname{}, but the dominant cost is wasted computation. Each block carries a boundary halo of neighbouring Gaussians, so primitives are duplicated across blocks and the same work is repeated on several GPUs, which surfaces directly as longer training time. Non-spatial sharding avoids this by giving every GPU one balanced, scene-wide slice (Appendix~\ref{sec:sharding_appendix}).

The system level is also ablated by sweeping the GPU count (Figure~\ref{fig:scalability}). Training time drops monotonically and nearly in parallel on both \emph{Rubble} and the larger \emph{Residence}, reaching $3.3$--$3.8\times$ speedup at eight GPUs while PSNR stays within a tight per-scene band. The scaling shape is thus largely independent of the number of input images, consistent with the system level workload model. Per-step compute is bounded by the per-GPU shard rather than by the number of training views. Appendix~\ref{sec:cost_appendix} gives a per-iteration cost breakdown, showing that the all-to-all stays under $7\%$ of iteration time and that the speedup is limited by fixed overhead rather than communication. This study runs on an 8$\times$A800 (80\,GB) node due to hardware availability; all other results use the default 4$\times$A6000 setup.

\section{Conclusion}
\label{sec:conclusion}

\methodname{} treats fast city-scale 3DGS as active workload control: every operation should touch as few Gaussians as possible without losing quality. It shards the model by index parity and distributes each render (\emph{system}), controls the population from both ends of densification while feeding importance back into density control (\emph{model}), and trims each camera's active set by LOD and importance (\emph{view}). With only an L1+SSIM loss and a scale regulariser, it matches recent large-scale baselines in quality while cutting training from hours to tens of minutes.

Limitations remain. The hard LOD gate can cause mild popping at level transitions, and aggressive simplification costs some perceptual fidelity on the texture-dense \emph{MatrixCity}, though not on the real aerial scenes. The importance schedule also uses fixed checkpoints rather than adapting per scene. Finally, \methodname{} assumes the global model fits the combined GPU memory. Scenes that exceed it would still need partition-based pipelines, which we leave to future work.

\bibliography{references}

\clearpage
\appendix

\section{3D Gaussian Splatting Preliminaries}
\label{sec:preliminaries}

3DGS~\citep{kerbl3Dgaussians} represents a scene with a set of explicit anisotropic primitives $\mathcal{G}=\{g_i\}_{i=1}^{N}$. Each Gaussian stores a center $\boldsymbol{\mu}_i$, a covariance $\boldsymbol{\Sigma}_i$, an opacity $\alpha_i$, and view-dependent color $\mathbf{c}_i(\mathbf{d})$ parameterised by spherical harmonics:
\begin{equation}
  G_i(\mathbf{x})=
  \exp\!\left(-\frac{1}{2}(\mathbf{x}-\boldsymbol{\mu}_i)^\top
  \boldsymbol{\Sigma}_i^{-1}(\mathbf{x}-\boldsymbol{\mu}_i)\right).
  \label{eq:gaussian}
\end{equation}
Following the standard 3DGS parameterization, the covariance is represented by learnable rotation and scale components.

For a camera view, Gaussians are projected to screen-space ellipses and rasterized by tile-based front-to-back alpha compositing. For pixel $\mathbf{p}$, the rendered color is
\begin{equation}
  \resizebox{\columnwidth}{!}{$\displaystyle
  \hat{\mathbf{C}}(\mathbf{p})=
  \sum_{i\in\mathcal{N}(\mathbf{p})}
  T_i(\mathbf{p})\,\alpha_i G'_i(\mathbf{p})\,\mathbf{c}_i(\mathbf{d}),
  \quad
  T_i(\mathbf{p})=\prod_{j<i}\left(1-\alpha_j G'_j(\mathbf{p})\right)
  $}
  \label{eq:alpha_compositing}
\end{equation}
where $\mathcal{N}(\mathbf{p})$ denotes the depth-ordered projected Gaussians overlapping the pixel, $G'_i$ is the projected 2D Gaussian value, and $T_i$ is accumulated transmittance. Training optimizes these primitive attributes from posed images and an initial point cloud, while adaptive density control periodically clones, splits, and prunes Gaussians. In city-scale reconstruction, the dominant cost is therefore not only the number of images, but also the number of active Gaussians that must be projected, exchanged across devices, rasterized, and updated at every iteration. \methodname{} is designed around reducing this active workload at the system, model, and view levels.

\section{Additional Implementation Details}
\label{sec:impl_appendix}

\paragraph{Optimization schedule.} Density control runs every $1{,}000$ steps from iteration $2{,}000$ up to $32{,}000$, with the last event at $31{,}000$, straddling the first importance-scoring pass ($T_1$); the LOD gate is disabled inside this window and engaged afterwards. After that pass, density-control gradients are reweighted by the per-Gaussian visibility factor $\phi_i$ for the remainder of the density-control window. The remaining implementation and hyperparameter details are in our code.

\paragraph{Baseline configurations.} All baselines are run on the same 4$\times$A6000 hardware, data, test splits, and evaluation resolution as \methodname{}. We evaluate the Grendel-GS baseline at two settings. \emph{Paper config} uses its authors' released training configuration. \emph{Long iter} extends the schedule---$500\mathrm{k}$ iterations on the Mill-19 and UrbanScene3D scenes and $1000\mathrm{k}$ on \emph{MatrixCity}, several times its default budget---to rule out under-training as the cause of its low quality. As Table~\ref{tab:main_results} shows, the extra iterations barely change quality (e.g., \emph{Building} $19.27{\to}19.50$~dB), indicating that Grendel-GS's deficit on these aerial scenes reflects its non-LOD, undistilled representation rather than an insufficient training budget.

\section{Primitive Sharding vs.\ Spatial Partitioning}
\label{sec:sharding_appendix}

\begin{figure}[t]
  \centering
  \includegraphics[width=0.98\columnwidth]{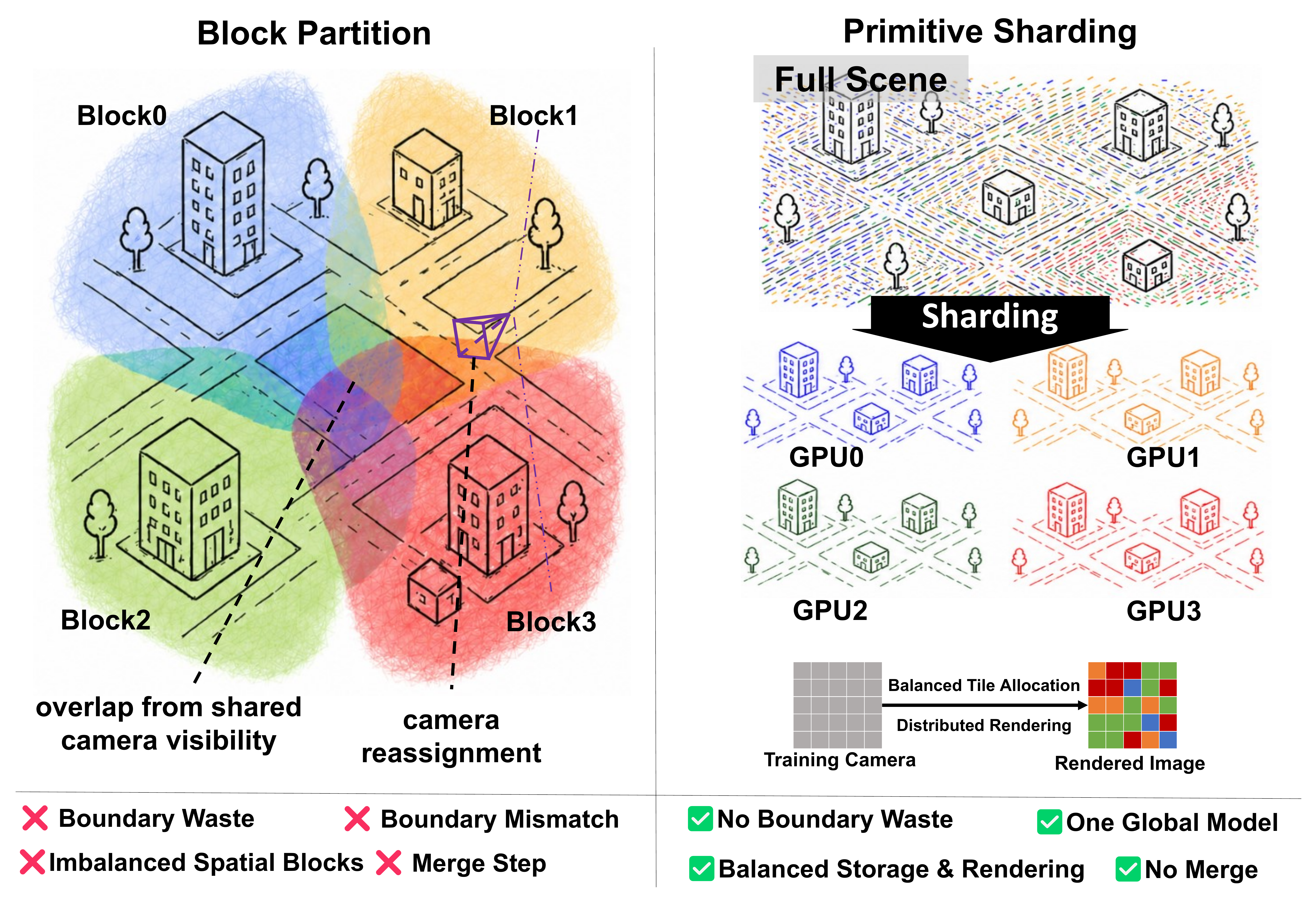}
  \caption{\textbf{Primitive sharding vs spatial block partitioning.} \emph{Left:} block partitioning binds each GPU to a fixed scene region, so cameras near boundaries trigger overlap and reassignment. \emph{Right:} \methodname{} shards one global Gaussian model across GPUs by index parity and distributes each render via balanced tile allocation, keeping both storage and rendering evenly spread.}
  \Description{A two-panel side-by-side comparison contrasting spatial block partitioning on the left with our index-parity primitive sharding on the right. The left panel depicts an aerial cityscape carved into four world-space blocks shown in distinct muted colors (blue, amber, green, red) that visibly overlap near block boundaries, with one training camera straddling several blocks via dashed sight lines, and four red cross marks at the bottom listing the inherent costs: cross-block visibility waste, boundary mismatch, imbalanced spatial blocks, and a costly final single-GPU merge step. The right panel shows the unified global Gaussian model at the top with all four colors interleaved across the entire scene, a thin labelled Sharding arrow leading downward to four per-GPU slices each rendered in its own single color, a small inset at the bottom showing how a training camera's image tiles are balanced-allocated across the four GPUs and composited into the final rendered image, and four green check marks at the bottom highlighting the corresponding wins: no boundary waste, one global model, balanced storage and rendering, and no merge step. Together the panels make a sharp visual case that non-spatial sharding plus tile-balanced distributed rendering eliminates the structural overheads of spatial decomposition.}
  \label{fig:sharding_vs_blocks}
\end{figure}

Figure~\ref{fig:sharding_vs_blocks} contrasts our index-parity primitive sharding with conventional spatial block partitioning. Block partitioning binds each GPU to a fixed world-space region: cameras near a block boundary see Gaussians owned by several GPUs, so the same view triggers cross-block visibility overlap and boundary reassignment, and a final single-GPU merge is needed to assemble the full scene. Because scene content and camera coverage are rarely uniform, the resulting shards are also imbalanced in both storage and compute. \methodname{} instead shards one global Gaussian model across GPUs by index parity, so every device holds a balanced scene-wide slice, and distributes each render step through balanced tile allocation. This removes the boundary waste, the imbalance, and the merge step, and keeps both storage and per-step rendering evenly spread as the GPU count grows.

Table~\ref{tab:sharding_cameras} quantifies this cost on \emph{Building} (1,920 training cameras, 4 GPUs), the scene used for the partition ablation in Table~\ref{tab:workload_ablation}. We split the scene into a uniform $4{\times}5$ ($20$) and $5{\times}6$ ($30$) grid, each block carrying a boundary halo of neighbouring Gaussians for cross-block consistency, and train the blocks independently across the four GPUs. Because camera views straddle block boundaries, each camera is assigned to $2.54$ blocks on average at $20$ blocks and $3.50$ at $30$, and $86.8\%$/$95.5\%$ of cameras fall in two or more blocks. These counts are geometric, set by the camera poses and block boundaries, so they hold for any spatial partition at this granularity, not just ours. Because each block also carries a boundary halo of neighbouring Gaussians, the overlapping blocks re-optimise the same surfaces, wasting computation that we quantify in time and model size below. \methodname{} instead trains one global model on a single camera stream, so each camera enters training once per step with no cross-block duplication. Its only distributed cost is the all-to-all exchange inside each render step, which Table~\ref{tab:cost_breakdown} shows stays under $7\%$ of iteration time.

This redundancy wastes work on two axes. In time, the equivalent wall-clock of Table~\ref{tab:workload_ablation} (block split, the four-GPU makespan of training all blocks, and the merge) reaches $115.6$ and $169.8$ min, about $3$ and $4\times$ \methodname{}'s $39$ min, dominated by the makespan. In model size, Table~\ref{tab:partition_gaussians} shows the deeper waste: the blocks collectively optimise $23.5$M ($20$ blocks) and $34.3$M ($30$ blocks) Gaussians, yet only $5.37$M and $5.43$M survive the merge. Most per-block Gaussians are redundant copies of the same surfaces produced by overlapping blocks, so $77\%$ and $84\%$ of the optimisation is discarded. Finer partitioning is worse on both axes. This is already the favourable case, with each block trained for only $20$k iterations. Training every block to a full schedule instead costs about $522$ min and $11.2$ h yet moves quality by at most $0.2$~dB, so the gap is structural redundancy, not an insufficient per-block budget.

\begin{table}[t]
  \centering
  \footnotesize
  \setlength{\tabcolsep}{4pt}
  \resizebox{\columnwidth}{!}{%
  \begin{tabular}{lccc}
    \toprule
     & \multicolumn{2}{c}{Spatial blocks} & \methodname{} \\
    \cmidrule(lr){2-3}
     & 20-block & 30-block & (sharding) \\
    \midrule
    Cameras covered (of 1,920)                & 1,898 & 1,912 & 1,920 \\
    Camera assignments                        & 4,884 (2.54$\times$) & 6,720 (3.50$\times$) & 1,920 (1.00$\times$) \\
    Cameras in $\geq$2 blocks                 & 1,667 (86.8\%) & 1,833 (95.5\%) & 0 (0\%) \\
    \bottomrule
  \end{tabular}%
  }
  \caption{\textbf{Camera coverage under spatial block partitioning vs.\ primitive sharding (\emph{Building}, 1,920 training cameras, 4 GPUs).} Both partitions use a uniform grid ($4{\times}5$ and $5{\times}6$) with boundary halos. Because views straddle block boundaries, each camera is claimed by several blocks; the assignment counts are geometric and hold for any spatial partition at this granularity. \methodname{} assigns every camera once.}
  \label{tab:sharding_cameras}
\end{table}

\begin{table}[t]
  \centering
  \footnotesize
  \setlength{\tabcolsep}{6pt}
  \begin{tabular}{lcc}
    \toprule
     & 20-block & 30-block \\
    \midrule
    Gaussians trained (sum over blocks) & 23.5M & 34.3M \\
    Merged final model                  & 5.37M & 5.43M \\
    Fraction surviving the merge        & 22.9\% & 15.8\% \\
    \bottomrule
  \end{tabular}
  \caption{\textbf{Gaussian survival under spatial block partitioning (\emph{Building}).} Each block optimises its own Gaussians, and the boundary halos make overlapping blocks model the same surfaces, so most per-block Gaussians are duplicates discarded at the merge. Finer partitioning wastes more: the 30-block split trains $46\%$ more Gaussians than the 20-block one yet keeps a smaller fraction. \methodname{} optimises a single global model (4.65M final) with no merge discard.}
  \label{tab:partition_gaussians}
\end{table}

\section{Per-Iteration Cost and Scaling Breakdown}
\label{sec:cost_appendix}

\begin{table}[t]
  \centering
  
  \footnotesize
  \setlength{\tabcolsep}{4pt}
  \resizebox{\columnwidth}{!}{%
  \begin{tabular}{c|cccccc|c}
    \toprule
    GPUs & Preproc. & Comm. & Render & Loss & Backward & Other & Total \\
    \midrule
    1 & 15.8 & --  & 23.2 & 13.2 & 113.6 & 20.2 & 186.9 \\
    2 & 9.5  & 3.7 & 13.6 & 6.9  & 67.4  & 17.6 & 119.7 \\
    3 & 7.9  & 4.2 & 11.8 & 5.4  & 49.3  & 16.8 & 96.3  \\
    4 & 8.3  & 3.2 & 7.9  & 3.5  & 40.5  & 14.9 & 79.2  \\
    5 & 9.3  & 3.7 & 6.9  & 3.1  & 35.4  & 16.6 & 75.9  \\
    6 & 10.6 & 4.2 & 6.5  & 2.8  & 31.0  & 17.5 & 73.6  \\
    7 & 11.4 & 4.5 & 5.9  & 2.5  & 28.1  & 16.3 & 69.7  \\
    8 & 13.0 & 4.4 & 5.6  & 2.3  & 25.8  & 16.5 & 68.6  \\
    \bottomrule
  \end{tabular}%
  }
  \caption{\textbf{Per-iteration cost breakdown (\emph{Residence}, A800 80\,GB).} Wall time in ms for the main training stages as the GPU count grows. \emph{Comm.}\ is the cross-GPU all-to-all; \emph{Other} pools per-rank data loading, the optimizer step, and bookkeeping. Components may not sum exactly to the total due to rounding and unlisted minor stages.}
  \label{tab:cost_breakdown}
\end{table}

\paragraph{Cost breakdown.} Table~\ref{tab:cost_breakdown} decomposes one training iteration on \emph{Residence} as the GPU count grows from one to eight, and explains how \methodname{} scales. First, \emph{communication is not the bottleneck}: the all-to-all that routes projected Gaussians to their tile owners costs only $3$--$5$\,ms per iteration---under $7\%$ of the total---and does not blow up with the GPU count, because each rank exchanges only the splats overlapping its own tiles rather than the whole scene. Second, the sub-linear end-to-end scaling stems from \emph{fixed overhead}, not from the sharded design: the compute-heavy backward pass scales near-linearly ($113.6\to25.8$\,ms from one to eight GPUs), while the per-rank data loading, optimizer, and bookkeeping (Other) stay roughly constant at $15$--$20$\,ms and come to dominate the shrinking iteration. This residual is addressable by GPU-side data decoding and is orthogonal to the workload-reduction mechanisms that are our focus.

\paragraph{Memory is sharded, not replicated.} Each rank stores only its $|\mathcal{G}|/M$ shard and its optimizer state, and during rendering receives only the projected Gaussians overlapping its own tiles---never a full copy of the scene. Per-rank peak memory therefore falls as the GPU count rises: on \emph{Rubble} (Figure~\ref{fig:scalability} setup) it drops from $17.2$\,GB at one GPU to $2.9$\,GB at eight. The binding limit is the \emph{aggregate} capacity across GPUs, as noted in our limitations, rather than any single-rank replication of the global model.

\begin{table}[t]
  \centering
  
  \footnotesize
  \setlength{\tabcolsep}{4pt}
  \resizebox{\columnwidth}{!}{%
  \begin{tabular}{lcc}
    \toprule
    Stage & \emph{Rubble} & \emph{MatrixCity} \\
    \midrule
    Octree initialization & 1.14M & 2.29M \\
    Before pass 1 ($T_1{=}16$k) & 5.22M & 7.07M \\
    After pass 1 & 3.65M ($-30\%$) & 4.69M ($-34\%$) \\
    Peak, end of densification (31k) & 7.38M & 13.90M \\
    Before pass 2 ($T_2{=}34$k) & 7.38M & 13.90M \\
    After pass 2 (final) & 4.36M ($-41\%$) & 6.91M ($-50\%$) \\
    \bottomrule
  \end{tabular}%
  }
  \caption{\textbf{Gaussian-count trajectory under the full pipeline.} Global Gaussian count at each stage of the schedule on \emph{Rubble} (90k) and \emph{MatrixCity} (120k). Percentages give the reduction of each simplification pass. The count is fixed after the second pass, and the last row equals the final model size.}
  \label{tab:gauss_trajectory}
\end{table}

\paragraph{Population trajectory.} Table~\ref{tab:gauss_trajectory} traces the global Gaussian count through the training schedule on \emph{Rubble} and \emph{MatrixCity}, with the spawn gate active as in all main results. Two regularities stand out. First, the passes reshape the model without capping the peak. The first pass removes $30$--$34\%$ of the population, yet densification rebuilds and then exceeds the pre-pass count within a few thousand iterations. Second, the population peaks at the last density-control event (iteration 31k) and holds that size until the second pass, which removes $41$--$50\%$ in a single step and fixes the final model. The transient peak, not the final size, sets per-GPU memory during training, and with the gate active it stays below the no-gate peaks of Table~\ref{tab:overshoot}, consistent with the peak-population cuts measured in Appendix~\ref{sec:oracle_appendix}.

\section{Gaussian Lifetime Measurements}
\label{sec:lineage_appendix}

This appendix reports the measurements that motivate and validate the spawn gate of Sec.~\ref{sec:natal}: how the population overshoots, which Gaussians die, and how predictable their survival is at spawn time.

\begin{figure}[t]
  \centering
  \includegraphics[width=0.98\columnwidth]{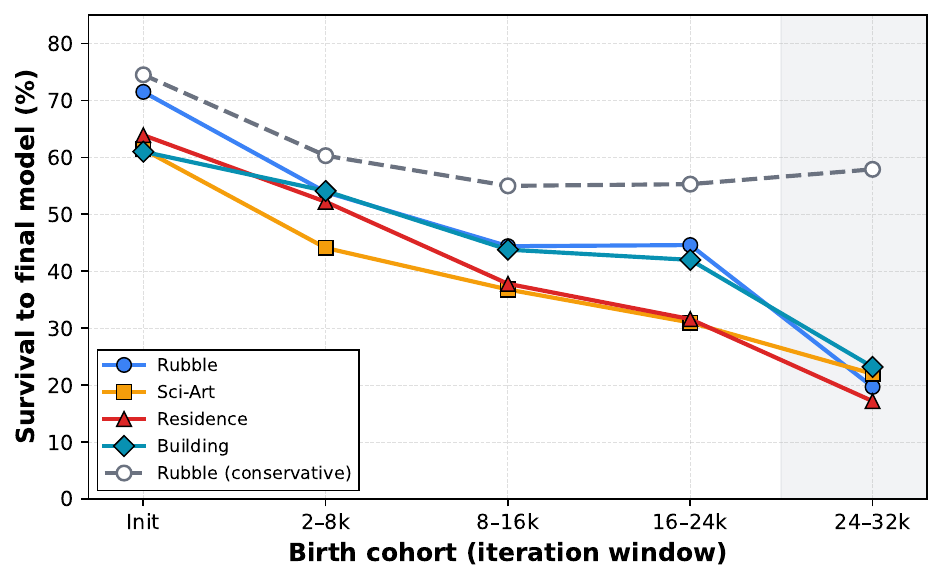}
  \caption{\textbf{Survival to the final model by birth cohort.} Each curve is
  the fraction of Gaussians born in a given iteration window that reach the
  final model. Under the aggressive densification schedule (four colored
  scenes) survival falls sharply with birth time, collapsing in the last
  densification window (shaded); under a conservative schedule (grey, dashed) it
  stays flat---the survival structure is schedule-induced.}
  \Description{Line plot of survival-to-final rate versus birth cohort for four
  scenes under the aggressive schedule, all declining sharply in the last
  densification window, plus a flat conservative-schedule control.}
  \label{fig:hazard}
\end{figure}

\paragraph{Population overshoot.} Resolving high-fidelity detail calls for an aggressive, low-threshold densification schedule that grows the Gaussian population large during training, after which scheduled simplification passes distil it into a compact final model. As a result, the population is at its largest mid-training rather than at the end. Table~\ref{tab:overshoot} quantifies this overshoot. The peak population reaches roughly twice the final size on both \emph{Rubble} and \emph{MatrixCity}. This transient peak, rather than the final size, dictates the per-GPU memory limits throughout the densification window, which is exactly why our spawn gate targets the peak rather than the final model.

\begin{table}[t]
  \centering
  
  \small
  \begin{tabular}{lccc}
    \toprule
    Scene & Peak & Final & Peak\,/\,Final \\
    \midrule
    \emph{Rubble}     & 8.55M  & 4.47M & $1.9\times$ \\
    \emph{MatrixCity} & 18.68M & 8.13M & $2.3\times$ \\
    \bottomrule
  \end{tabular}
  \caption{\textbf{Population overshoot (no spawn gate).} Live Gaussian count at the mid-training peak versus the final model; the peak is $1.9$--$2.3\times$ larger.}
  \label{tab:overshoot}
\end{table}

\paragraph{Lineage instrumentation.} To measure per-Gaussian lifetimes exactly, we assign every Gaussian a persistent identifier at creation. We log each birth with its spawn-time features and parent link, record each death with the mechanism that removed it, and carry identifiers through the redistribution all-to-all so lineage survives resharding. The record is exact, reconciling logged births and deaths with the live population at every checkpoint (10.25M births and 5.78M deaths on \emph{Rubble}), and adds no measurable training time. Lineage runs are used only offline to fit the gate; no logging runs during the gated training reported in this paper.

\paragraph{Survival structure.} Survival to the final model falls with birth time (Fig.~\ref{fig:hazard}): initialization points survive at $71.5\%$, while only $17$--$23\%$ of Gaussians born in the last densification window survive, consistently across all four scenes we instrumented. \emph{Rubble}, retrained under a conservative schedule (higher densification threshold), shows flat survival of $55$--$60\%$ across cohorts, so the structure is induced by the aggressive schedule---exactly the regime that wins quality. A delayed-judgment control shows the effect is not merely that late spawns are examined young: moving the second simplification pass from iteration 34k to 50k, so the late cohort matures before judgment, raises its survival from $19.7\%$ to $35.9\%$, but at matched age late-born Gaussians still survive $9$ points less often than earlier cohorts. Late birth is penalized twice: late spawns are intrinsically weaker at matched age, and younger when examined.

\paragraph{Spawn gate features and veto.} For a spawn candidate with parent $j$ at densification step $t$, we assemble the spawn-time feature vector
\begin{equation}
  \mathbf{x}_{j,t}=\bigl(\bar g_j,\ \alpha_j,\ \bar s_j,\ \tau_j,\ d_j,\ \ell_j,\ t,\ \mathbf{1}[\mathrm{split}]\bigr),
\end{equation}
where $\bar g_j$ is the view-space positional gradient magnitude averaged over the $d_j$ views that observed the parent (the same signal the densification threshold tests), $\alpha_j$ the parent opacity, $\bar s_j$ the mean of the parent's three scale axes, $\tau_j$ the parent's age in iterations since its own birth, $d_j$ the densification denominator (the number of views that accumulated gradient onto the parent), and $\ell_j$ the child's LOD level---the parent's level for a clone, one level finer for a split. The gradient-boosted tree predicts survival $p_j=f(\mathbf{x}_{j,t})\in[0,1]$. Clones and splits form separate candidate sets at each step; within a set $\mathcal{C}_t$ the gate vetoes the $\lceil\rho\lvert\mathcal{C}_t\rvert\rceil$ candidates of lowest predicted survival,
\begin{equation}
  \mathcal{V}_t=\!\!\operatorname*{arg\,min}_{\substack{S\subseteq\mathcal{C}_t\\ \lvert S\rvert=\lceil\rho\lvert\mathcal{C}_t\rvert\rceil}}\ \sum_{j\in S}p_j,\qquad \mathcal{C}'_t=\mathcal{C}_t\setminus\mathcal{V}_t,
\end{equation}
with $\rho=0.6$; only $\mathcal{C}'_t$ is instantiated. Vetoing by rank within $\mathcal{C}_t$, rather than thresholding $p_j$ at a fixed value, makes the gate depend only on the ordering of $f$, so a single predictor transfers across scenes despite large differences in candidate volume and score calibration.

\begin{figure}[t]
  \centering
  \includegraphics[width=0.98\columnwidth]{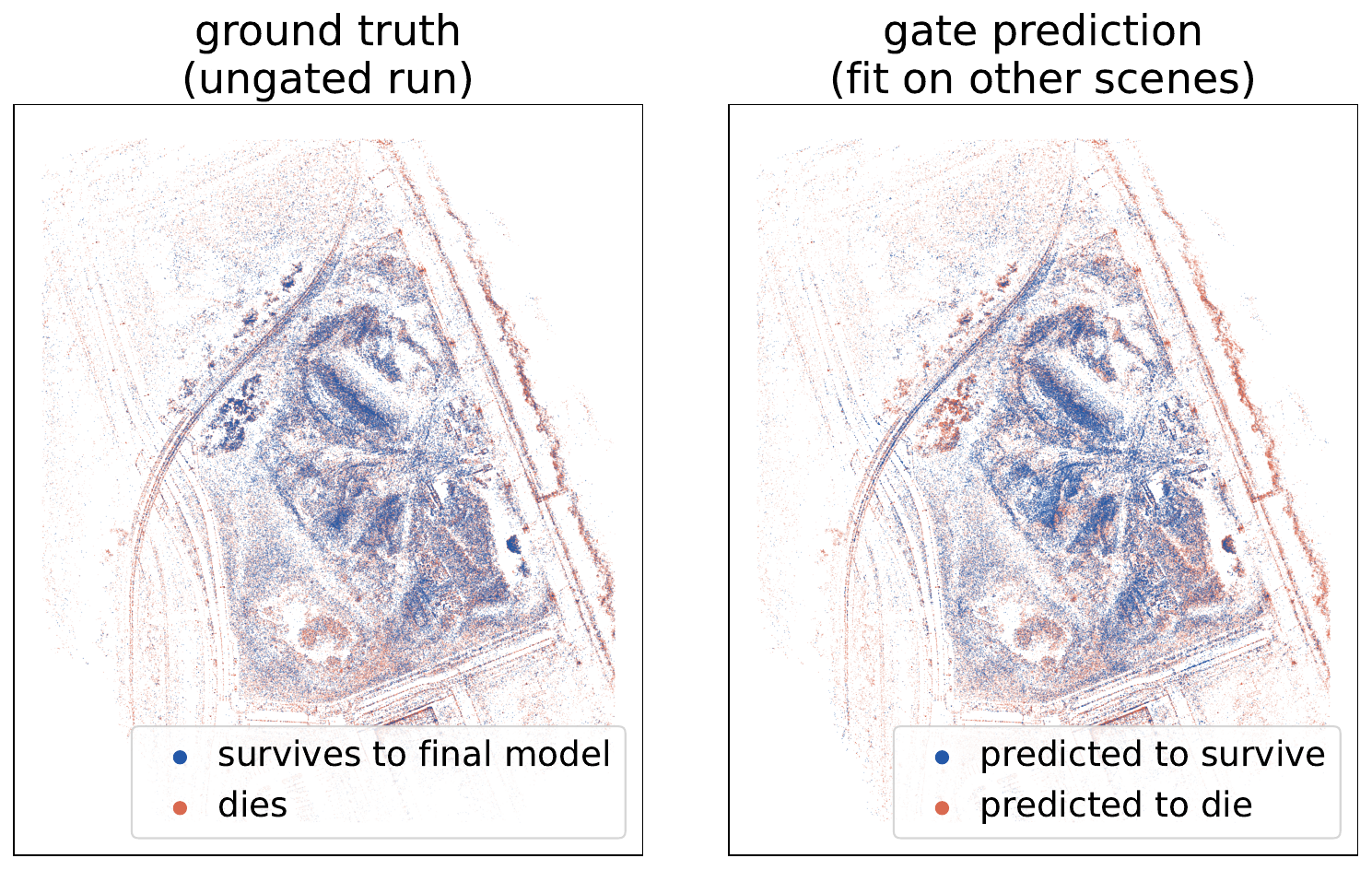}
  \caption{\textbf{Ground-truth survival vs.\ gate prediction (\emph{Rubble}).} Each point is a spawn born between iterations 24k and 30k of an ungated, lineage-instrumented run, plotted at its scene location and colored by whether it survives to the final model: \emph{left,} the actual outcome; \emph{right,} the spawn-time prediction of the leave-one-scene-out gate, rate-matched to the true death rate so the panels differ only in where the colors fall. Per-region predicted and actual survival rates correlate at $r{=}0.84$, although the predictor uses no positional features and never saw \emph{Rubble}.}
  \Description{Two side-by-side top-down scatter plots of the Rubble scene showing Gaussians born between iterations 24k and 30k of an ungated run. The left panel colors each point by its actual outcome, blue for survives to the final model and orange for dies; survivors trace the dense central rubble mound while deaths concentrate along the scene periphery, the roads, and sparsely observed border regions. The right panel colors the same points by the survival predicted at birth by the leave-one-scene-out spawn gate, replayed offline and rate-matched to the true death rate so both panels contain identical color proportions; the predicted-survivor and predicted-death regions closely mirror the left panel, showing that a predictor with no positional features, fit without ever seeing Rubble, reproduces the scene's spatial mortality structure.}
  \label{fig:natal_geometry}
\end{figure}

\paragraph{Predictability at birth.} The spawn gate is only viable if a candidate's survival is predictable from the features available at spawn time; Table~\ref{tab:natal_auc} shows that it is, and robustly so. When each logged spawn is labelled by whether it survives to the final model, the gradient-boosted tree predictor of Sec.~\ref{sec:natal} separates survivors from non-survivors at a held-out AUC of $0.748$--$0.787$. This holds on every scene and schedule we instrumented, including the conservative schedule and the delayed-judgment control, so the signal is not an artifact of one configuration. The strongest single spawn feature, parent opacity, reaches an AUC of only $0.63$--$0.66$ alone, so the predictability is genuinely multivariate. Recall also that every experiment in Sec.~\ref{sec:experiments} deploys a predictor fit exclusively on the \emph{other} scenes' lineage (Sec.~\ref{sec:natal}), and the predictability survives this restriction. Table~\ref{tab:gate_protocol} lists each deployed predictor's training pool and its transfer. Evaluated on the spawns of the scene it never saw, each leave-one-scene-out predictor reaches an AUC of $0.750$--$0.770$, only $0.013$--$0.021$ below the predictor fit on that scene's own lineage (Table~\ref{tab:natal_auc}). Transfer is harder across the domain gap to the synthetic \emph{MatrixCity}, where the deployed predictor reaches $0.702$; an ungated lineage run of that scene nonetheless lands inside the gated replicate band ($26.882$ vs.\ $26.877\pm0.021$), so even this weaker ordering leaves quality intact at the deployed rate. Together these establish the gate's premise: survival is predictable at birth, the prediction requires learning, and it transfers to unseen scenes. Figure~\ref{fig:natal_geometry} shows the transfer spatially. Replayed on a scene it never saw, the leave-one-scene-out predictor reproduces the spatial structure of ground-truth survival (per-region survival rates correlate at $r{=}0.84$) despite having no positional features.

\begin{table}[t]
  \centering
  
  \footnotesize
  \setlength{\tabcolsep}{3pt}
  \begin{tabular}{llcccc}
    \toprule
    &&&& \multicolumn{2}{c}{AUC} \\
    \cmidrule(lr){5-6}
    Scene & Schedule & Spawns & Surv. & full & opacity \\
    \midrule
    \emph{Rubble}     & aggressive        & 9.1M  & $40\%$ & $\mathbf{0.780}$ & $0.643$ \\
    \emph{Sci-Art}    & aggressive        & 3.8M  & $34\%$ & $\mathbf{0.787}$ & $0.634$ \\
    \emph{Residence}  & aggressive        & 8.4M  & $32\%$ & $\mathbf{0.764}$ & $0.636$ \\
    \emph{Building}   & aggressive        & 11.5M & $38\%$ & $\mathbf{0.772}$ & $0.657$ \\
    \midrule
    \emph{Rubble}     & conservative      & 2.3M  & $57\%$ & $\mathbf{0.768}$ & $0.643$ \\
    \emph{Rubble}     & short window      & 7.0M  & $47\%$ & $\mathbf{0.748}$ & $0.662$ \\
    \emph{Rubble}     & delayed judgment  & 9.1M  & $46\%$ & $\mathbf{0.763}$ & $0.648$ \\
    \bottomrule
  \end{tabular}
  \caption{\textbf{Spawn-time survival predictability.} Held-out AUC on each lineage-instrumented run: \emph{full} is the gradient-boosted predictor on all spawn-time features; \emph{opacity} uses the strongest single feature (parent opacity) alone as the ranking score, on the same held-out spawns. Surv.\ is the fraction of all logged spawns surviving to the final model (all birth cohorts pooled, initialization points excluded; per-cohort rates in Fig.~\ref{fig:hazard}). \emph{Conservative} raises the densification threshold, \emph{short window} ends densification at 24k instead of 32k, and \emph{delayed judgment} moves the second simplification pass from 34k to 50k.}
  \label{tab:natal_auc}
\end{table}

\begin{table}[t]
  \centering
  
  \footnotesize
  \setlength{\tabcolsep}{4pt}
 \begin{tabular}{llc}
    \toprule
    Eval.\ scene & Gate fit on & AUC \\
    \midrule
    \emph{Rubble}     & \emph{Sci-Art}+\emph{Residence}+\emph{Building} & $0.759$ \\
    \emph{Sci-Art}    & \emph{Rubble}+\emph{Residence}+\emph{Building}  & $0.770$ \\
    \emph{Residence}  & \emph{Rubble}+\emph{Sci-Art}+\emph{Building}    & $0.750$ \\
    \emph{Building}   & \emph{Rubble}+\emph{Sci-Art}+\emph{Residence}   & $0.759$ \\
    \emph{MatrixCity} & all four scenes                                 & $0.702$ \\
    \bottomrule
  \end{tabular}
  \caption{\textbf{Deployed gate protocol and transfer.} Each evaluation scene is gated by a predictor that never saw its own lineage; \emph{MatrixCity} uses a predictor fit on all four instrumented scenes. \emph{AUC} is the deployed predictor scored on all spawns of a lineage-instrumented run of the evaluation scene; the \emph{MatrixCity} lineage run serves evaluation only and enters no training pool.}
  \label{tab:gate_protocol}
\end{table}

\paragraph{Design rationale.}
The measurements above dictate four choices in the gate design (Sec.~\ref{sec:natal}). \emph{Why learned rather than heuristic?} The survival structure is schedule-induced (Survival structure), so a hand-crafted rule tuned to one regime breaks when the schedule changes, and single spawn signals such as parent opacity separate far more weakly (Predictability at birth). The ranking must be survival-correlated so that the vetoed fraction falls on non-surviving spawns rather than on true survivors, and the learned predictor supplies that correlation at a transferred AUC of $0.750$--$0.770$ on scenes it never saw (Table~\ref{tab:gate_protocol}); the oracle replay confirms that vetoes chosen this way carry no measurable end-to-end quality cost (Appendix~\ref{sec:oracle_appendix}). A new schedule does require a new predictor, but the adaptation is mechanical (collect lineage under the new schedule and refit), whereas a heuristic must be re-derived by hand for every regime. \emph{Why a rank veto rather than a probability threshold?} The gate's deliverable is a deterministic peak-population cut, and vetoing by rank fixes the veto rate at exactly $\rho$ on every scene by construction. A fixed threshold instead makes the rate an uncontrolled, scene-dependent quantity: a well-calibrated score must track each scene's base survival rate, and these rates differ across scenes (Table~\ref{tab:natal_auc}, Surv.\ column), so the same cutoff removes a different fraction of candidates on each. The leave-one-scene-out results confirm exactly the property the rank veto relies on. AUC is invariant to any monotone rescaling of the scores, so it measures the transfer of the ordering and nothing else (Predictability at birth). If per-scene rates prove desirable, they should enter as a deliberately chosen $\rho$ (Appendix~\ref{sec:sensitivity_appendix}), not as an uncontrolled by-product of calibration. \emph{Why the late window only?} Survival drops sharply after iteration 24k (Survival structure), and the sensitivity analysis confirms the optimal onset sits at this drop (Appendix~\ref{sec:sensitivity_appendix}). \emph{Why decide at spawn time rather than later?} A later decision would be better informed, but it forfeits the deliverable. Delaying the veto by $\Delta$ iterations leaves each non-surviving spawn allocated for those $\Delta$ iterations, so the peak-population cut shrinks as the delay grows. Waiting all the way to the scheduled simplification passes recovers no peak at all, and that endpoint is precisely the reactive removal the pipeline already provides. Only a spawn-time veto converts a prediction into memory that is never allocated (Appendix~\ref{sec:sensitivity_appendix}). It also saves the vetoed candidates' optimization, communication, and rasterization costs.

\subsection{Sensitivity to the Gate Rate and Onset}
\label{sec:sensitivity_appendix}

The gate exposes two hyperparameters: the veto fraction $\rho$ and the onset iteration at which it activates. We vary each on \emph{Building} under the aggressive headline schedule, holding the other fixed and changing nothing else (Table~\ref{tab:gate_sensitivity}). The $\rho{=}0.6$, onset~$24\mathrm{k}$ row is the deployed configuration, common to both panels.

\begin{table}[t]
\centering

\footnotesize
\setlength{\tabcolsep}{3.8pt}
\begin{tabular}{lccccc}
\toprule
        & PSNR  & Peak   & P.\,mem & A.\,mem & Time \\
        & (dB)  & (M)    & (GB)    & (GB)    &      \\
\midrule
\multicolumn{6}{l}{\emph{Veto rate} $\rho$ \emph{(onset }$24\mathrm{k}$\emph{)}} \\
$0$ (no gate) & 23.40 & 10.89 & 8.2 & 2.2 & 40m47s \\
$0.30$        & 23.42 & 9.79 & 7.3 & 2.1 & 39m59s \\
$0.45$        & 23.46 & 9.23 & 7.0 & 2.1 & 39m55s \\
$\mathbf{0.60}$ & \textbf{23.46} & \textbf{8.66} & \textbf{6.6} & \textbf{2.0} & \textbf{39m15s} \\
$0.75$        & 23.34 & 8.12 & 6.2 & 1.9 & 38m23s \\
$0.90$        & 23.24 & 7.57 & 5.8 & 1.8 & 37m41s \\
\midrule
\multicolumn{6}{l}{\emph{Onset (rate }$\rho{=}0.6$\emph{)}} \\
$16\mathrm{k}$ & 23.11 & 6.24 & 4.7 & 1.6 & 35m49s \\
$20\mathrm{k}$ & 23.25 & 7.49 & 5.7 & 1.8 & 37m20s \\
$\mathbf{24\mathrm{k}}$ & \textbf{23.46} & \textbf{8.66} & \textbf{6.6} & \textbf{2.0} & \textbf{39m15s} \\
$28\mathrm{k}$ & 23.28 & 9.80 & 7.5 & 2.1 & 39m59s \\
\bottomrule
\end{tabular}
  \caption{\textbf{Gate hyperparameter ablation on \emph{Building} (90k, aggressive schedule).} \emph{Top:} veto rate $\rho$ with onset fixed at $24\mathrm{k}$. \emph{Bottom:} onset with $\rho$ fixed at $0.6$. \emph{Peak} is the mid-training live Gaussian count; \emph{P.\,mem} and \emph{A.\,mem} are per-GPU peak and mean allocated memory. The bold row is the deployed configuration, common to both panels.}
  \label{tab:gate_sensitivity}
\end{table}

\paragraph{The gate is a peak-memory lever.} Per-GPU peak memory falls monotonically with $\rho$ and tracks the peak population almost linearly, while training time falls only $2$--$8\%$ even as the peak falls $10$--$30\%$: the reclaimed resource is the transient mid-training population peak, not the sustained per-iteration cost, consistent with the overshoot analysis (Appendix~\ref{sec:lineage_appendix}).

\paragraph{Quality is flat through the deployed rate; onset is fixed by the survival structure.} All veto rates $\rho{\le}0.6$ match the ungated quality, so vetoing up to $60\%$ of late candidates is quality-neutral; a measurable decline appears only at $\rho{\ge}0.75$. We therefore adopt $\rho{=}0.6$, the largest reclaim inside the flat region. The onset was fixed in advance at the sharp late-window drop in survival (Survival structure), and the sensitivity analysis confirms it: activating at $16\mathrm{k}$ begins gating before the drop (Fig.~\ref{fig:hazard}) and loses $0.35$\,dB, while $28\mathrm{k}$ gains nothing in quality yet returns the peak to $9.80$M, nearly the ungated $10.89$M. The deployed $24\mathrm{k}$ is the latest onset that still suppresses the overshoot.
\section{Oracle-Ceiling Analysis of the Spawn Gate}
\label{sec:oracle_appendix}

This appendix measures the deployed gate against the ceiling of what any spawn-time veto can achieve, rather than against the floor of not gating at all. The component ablation (Table~\ref{tab:workload_ablation}) establishes that floor, but cannot say how far the learned predictor sits from the best possible veto, or whether its mistakes, the true survivors it discards, cost anything end to end. As an empirical ceiling, we construct a perfect-foresight \emph{oracle} veto, with exact knowledge of each spawn's survival outcome in a lookahead run (exact labels, incomplete coverage; replay protocol below), and ask how much of that ceiling the deployed predictor captures. The answer (Table~\ref{tab:oracle}) is that the learned gate matches the ceiling's quality and captures $98\%$ of its peak-population cut at the deployed rate.

\paragraph{Replay protocol.} Multi-GPU training is non-deterministic (rasterization atomics), so the same parent may clone in one run, split in another, or not densify at all, and a naive oracle is ill-defined. All arms therefore branch from a single checkpoint at iteration 24k with identical populations and identical persistent identifiers. A lookahead branch trains to 90k un-gated with lineage logging and labels every spawn event in the gate window (the iterations where the spawn gate is active, $[24\mathrm{k},32\mathrm{k})$; Sec.~\ref{sec:natal}) by the key (parent identifier, densification iteration, clone or split): an event is \emph{non-surviving} if and only if none of its children survives to the final model; in the lookahead run, $78.8\%$ of the 2.74M spawn events in the gate window are non-surviving. The oracle arms replay from the same checkpoint, look up each candidate's key, and veto only candidates labeled non-surviving; unrecognized candidates, whose exact event never occurred in the lookahead because the trajectories drift after the branch, are always kept. The measured match rate is $100\%$ at the branch point, $65\%$ by the window's end, and $78\%$ over all candidates in the gate window. Since unrecognized candidates are always kept, this coverage gap can only make the measured ceiling conservative, mildly understating a true oracle's, so it does not affect the ceiling comparison. A per-cohort cap yields an oracle arm rate-matched to the deployed veto fraction $\rho{=}0.6$.

\begin{table}[t]
\centering

\footnotesize
\setlength{\tabcolsep}{3.5pt}
\resizebox{\columnwidth}{!}{%
\begin{tabular}{lcccc}
\toprule
Arm & PSNR$\uparrow$ & $\Delta$PSNR & Peak pop. & $\Delta$peak \\
\midrule
Clean reference (no gate)           & 27.731 & ---     & 9.22M & --- \\
Oracle, uncapped                    & 27.821 & $+0.09$ & 7.65M & $-17.0\%$ \\
Oracle, rate-matched ($\rho{=}0.6$) & 27.704 & $-0.03$ & 7.70M & $-16.5\%$ \\
Learned gate (LOSO, $\rho{=}0.6$)   & 27.821 & $+0.09$ & 7.69M & $-16.6\%$ \\
\midrule
Random veto ($\rho{=}0.6$)          & 27.604 & $-0.13$ & 7.69M & $-16.6\%$ \\
\bottomrule
\end{tabular}%
}
  \caption{\textbf{Oracle ceiling and random floor on \emph{Rubble} (90k).} All arms resume from the same 24k checkpoint; numbers are comparable within this table only, not against the fresh-run tables in the main paper. The last row replaces the learned ranking with a random veto at the same rate and onset.}
  \label{tab:oracle}
\end{table}

\paragraph{Findings.} The table says four things. First, the learned gate matches the oracle's quality and captures $98\%$ of its peak-population cut, so better prediction has little room left at the deployed rate. Second, every informed veto arm, from perfect foresight to the learned gate, reaches the quality of ungated training, so the gate's deliverable is the peak-population cut at no quality cost. Third, the rate-matched oracle, which never wrongly vetoes a survivor, is on par with the learned gate, so the gate's misclassifications carry no visible cost. Fourth, the random veto sits clearly at the bottom, $0.22$~dB below the learned gate, the one large gap in the column: an uninformed cut at the same rate and onset costs real quality. The gate is therefore quality-neutral not because vetoing this many candidates is inherently safe, but because its ranking concentrates the vetoes on spawns that would have died anyway. What the ranking demonstrably changes is which candidates are removed. A random veto discards true survivors at the $21.2\%$ base rate. The learned veto, replayed offline on the same lookahead events with the transferred predictor of Table~\ref{tab:gate_protocol}, discards them at $11.1\%$, roughly half as many wrongly vetoed survivors at the same rate. On scenes where more candidates survive, or at higher veto rates, an uninformed veto discards correspondingly more true survivors, so this deficit is a floor rather than a worst case. The robust claims are the label statistics, the peak-population cuts that are stable across arms, the parity in quality between the informed arms and ungated training, and the random floor.

\section{Additional Qualitative Comparisons}
\label{sec:qualitative_appendix}

\begin{figure*}[p]
  \centering
  \includegraphics[width=\textwidth,height=0.94\textheight,keepaspectratio]{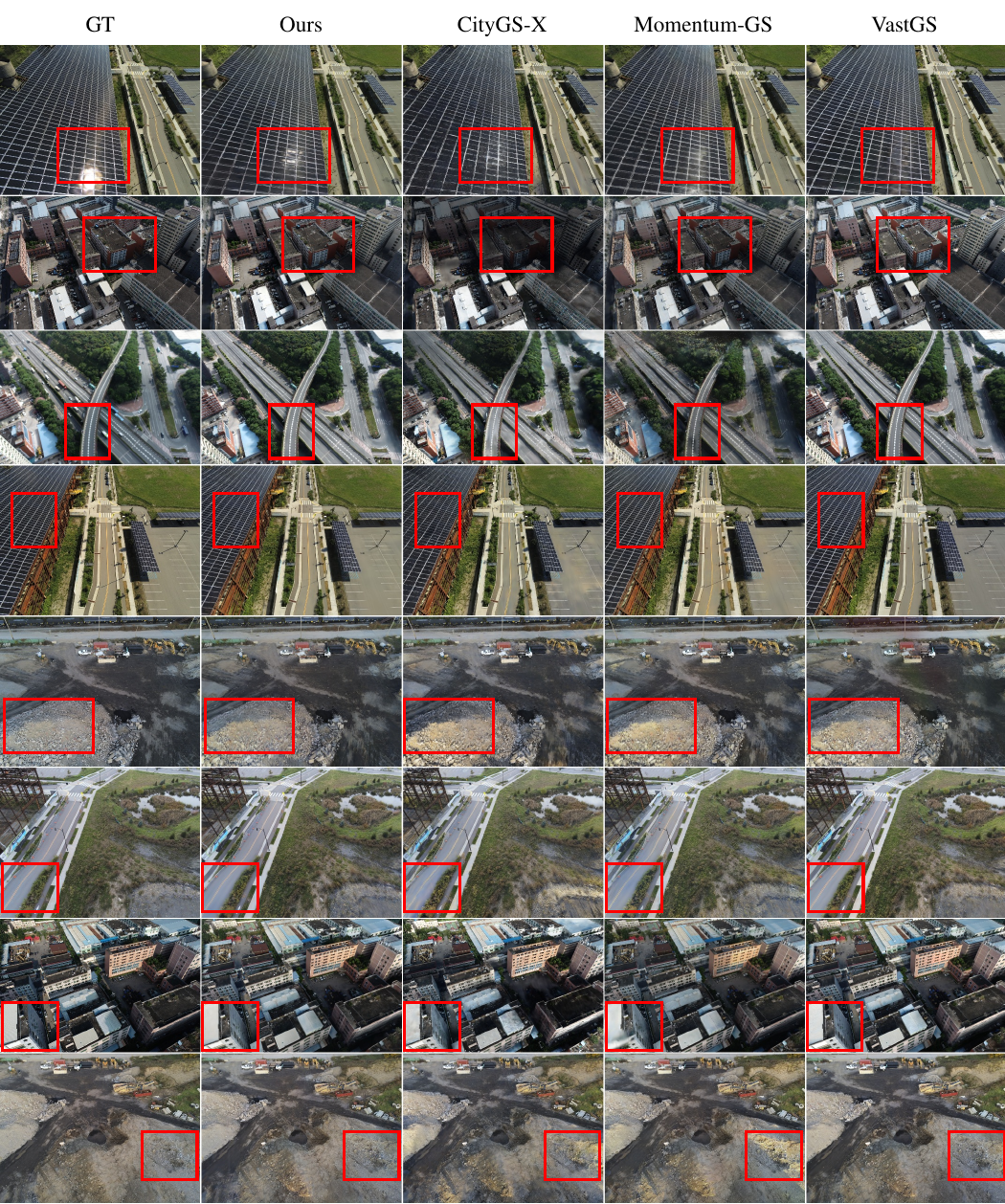}
  \caption{\textbf{Qualitative comparison on city-scale scenes.} Eight held-out test views spanning the benchmark scenes; columns show GT, \methodname{} (Ours), CityGS-X, Momentum-GS, and VastGaussian. Red rectangles highlight close-ups where \methodname{} preserves detail comparable to or sharper than the baselines.}
  \Description{A full-page qualitative comparison grid of eight held-out test views spanning diverse large-scale aerial benchmark scenes, organised as a clean five-column by eight-row matrix. The columns are labelled, from left to right, GT (ground truth), Ours (BlitzGS), CityGS-X, Momentum-GS, and VastGaussian. Each row shows one held-out test view, allowing strict side-by-side comparison across methods. Vivid crimson rectangles consistently highlight zoomed-in close-up regions, including building silhouettes, facade textures, road markings, vegetation foliage, and shadow boundaries, drawing the reader's eye to fine detail recovery. Across the eight rows, BlitzGS renderings recover comparable structural detail and texture sharpness to the strongest baseline on most views, with only mild popping artifacts visible at LOD transitions under extreme close-ups, vividly demonstrating that the order-of-magnitude training-time reduction is achieved without sacrificing perceptual fidelity.}
  \label{fig:visual_comparison}
\end{figure*}

\begin{figure*}[p]
  \centering
  \includegraphics[width=\textwidth,height=0.94\textheight,keepaspectratio]{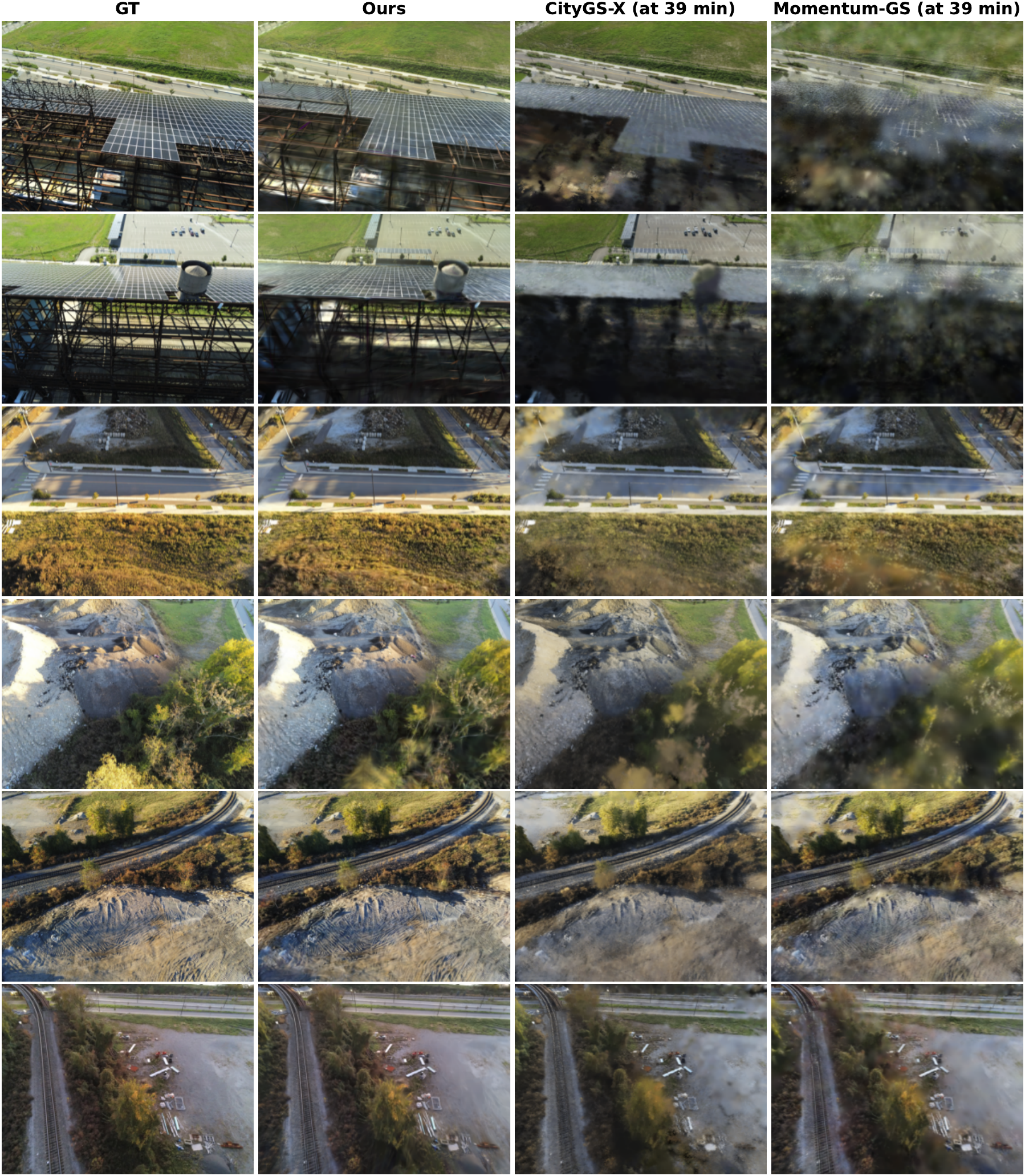}
  \caption{\textbf{Qualitative comparison at the same training budget ($\approx 39$ minutes).} Each baseline is shown at its checkpoint nearest \methodname{}'s wall-clock budget; at this point CityGS-X and Momentum-GS are still under-trained while \methodname{} has finished.}
  \label{fig:visual_at_35min}
\end{figure*}

Figure~\ref{fig:visual_comparison} compares \methodname{} against recent large-scale baselines on eight held-out test views spanning our benchmark scenes, with the ground truth alongside for reference. Figure~\ref{fig:visual_at_35min} instead fixes the training budget: every baseline is shown at the checkpoint closest to \methodname{}'s wall-clock time ($\approx 39$ minutes), which is the point at which \methodname{} has already finished training.


\end{document}